\documentclass[%
superscriptaddress,
showpacs,
showkeys,
preprintnumbers,
 amsmath,amssymb,
 aps,
prd, twocolumn, 
floatfix,
usenames,
dvipsnames
]{revtex4-1}

\usepackage{epsfig}
\usepackage{graphicx}
\usepackage{dcolumn}
\usepackage{bm}
\usepackage[colorlinks=true,linkcolor=blue,citecolor=blue]{hyperref}
\usepackage{enumitem}

\usepackage{natbib}
\usepackage{comment}
\usepackage{xcolor}
\usepackage{enumitem}
\usepackage{makecell, boldline}
\usepackage{gensymb}
\usepackage[normalem]{ulem}
\usepackage{subfigure}
\usepackage{multirow}
\usepackage{booktabs}
\usepackage{tabularx}

\newcommand{\yr}{\, \text{yr}}
\newcommand{\degsq}{\, \rm deg^2}

\newcommand{\SNR}{\text{SNR}}

\newcommand{\mchirp}{\mathcal{M}}

\newcommand{\msun}{{\, \rm M}_\odot}

\newcommand{\obsc}{obscuration }

\newcommand\mnras{Monthly Notices of the Royal Astronomical Society}
\newcommand\apjl{The Astrophysical Journal Letters}

\newcommand\pasa{Publications of the Astronomical Society of Australia}
\newcommand\jcap{Journal of Cosmology and Astroparticle Physics}
\newcommand\aj{Astronomical Journal}
\newcommand\aapr{The Astronomy and Astrophysics Review}
\newcommand{\aap}{Astronomy\&Astrophysics}
\newcommand{\araa}{Annual Review of Astronomy and Astrophysics}
\newcommand{\ssr}{Space Science Reviews}
\newcommand\Tstrut{\rule{0pt}{2.4ex}}       
\newcommand\Bstrut{\rule[-1.3ex]{0pt}{0pt}} 

\begin{document}

\preprint{APS/123-QED}

\title{Massive black hole binaries in LISA: multimessenger prospects and electromagnetic counterparts}

\author{Alberto Mangiagli}
\email{mangiagli@apc.in2p3.fr}
\affiliation{Universit\'e Paris Cit\'e, CNRS, Astroparticule et Cosmologie, F-75013 Paris, France}

\author{Chiara Caprini}
\affiliation{Universit\'e de Gen\'eve, D\'epartement de Physique Th\'eorique and Centre for Astroparticle Physics,
24 quai Ernest-Ansermet, CH-1211 Gen\'eve 4, Switzerland}
\affiliation{CERN, Theoretical Physics Department, 1 Esplanade des Particules, CH-1211 Gen\'eve 23, Switzerland}

\author{Marta Volonteri}
\affiliation{Institut d’Astrophysique de Paris, CNRS \& Sorbonne Universit\'e, UMR 7095, 98 bis bd Arago, F-75014 Paris, France}

\author{Sylvain Marsat}
\affiliation{Laboratoire des 2 Infinis - Toulouse (L2IT-IN2P3), Universit\'e de Toulouse, CNRS, UPS, F-31062 Toulouse Cedex 9, France}

\author{Susanna Vergani}
\affiliation{LUTH, GEPI and LERMA, Observatoire de paris, CNRS, PSL University, 5 place Jules Janssen, 92190,
Meudon, France}

\author{Nicola Tamanini}
\affiliation{Laboratoire des 2 Infinis - Toulouse (L2IT-IN2P3), Universit\'e de Toulouse, CNRS, UPS, F-31062 Toulouse Cedex 9, France}

\author{Henri Inchausp\'e}
\affiliation{Universit\'e Paris Cit\'e, CNRS, Astroparticule et Cosmologie, F-75013 Paris, France}
\affiliation{Institut f\"ur Theoretische Physik, Universit\"at Heidelberg, Philosophenweg 16, 69120 Heidelberg, Germany}

\date{\today}

\begin{abstract}
In the next decade, the Laser Interferometer Space Antenna (LISA) will detect the coalescence of massive black hole binaries (MBHBs) in the range $[10^4, 10^8] \, \msun$,  up to $z\sim10$. Their gravitational wave (GW) signal is expected to be accompanied by an electromagnetic counterpart (EMcp), generated by the gas accreting on the binary or on the remnant BH.
In this work, we present the number and characteristics (such as redshift and mass distribution, apparent magnitudes or fluxes) of EMcps detectable jointly by LISA and some representative EM telescopes.
We combine state-of-the-art astrophysical models for the galaxies formation and evolution to build the MBHBs catalogues, with Bayesian tools to estimate the binary sky  position uncertainty from the GW signal. 
Exploiting additional information from the astrophysical models, such as the amount of accreted gas and the BH spins, we evaluate the expected EM emission in the soft X-ray, optical and radio bands.
Overall, we predict between 7 and 20 EMcps in 4 yrs of joint observations by LISA and the considered EM facilities, depending on the astrophysical model.
We also explore the impact of the hydrogen and dust obscuration of the optical and X-ray emissions, as well as of the collimation of the radio emission: 
these effects reduce the number to EMcps to 2 or 3, depending on the astrophysical model, again in 4 yrs of observations.
Most of the EMcps are characterised by faint EM emission, challenging the observational capabilities of future telescopes.
Finally, we also find that systems with multi-modal sky position posterior distributions represent only a minority of cases and do not affect significantly the number of EMcps. 

\end{abstract}


\maketitle


\section{\label{sec:intro}Introduction}

The Laser Interferometer Space Antenna (LISA)\cite{Seoane17}  is planned for launch in 2034 and will detect gravitational waves (GWs) in between $[10^{-4}, 10^{-1}] \, \rm Hz$.  Among other sources, LISA will detect the coalescence of massive black hole binaries (MBHBs) in the entire Universe up to redshift $z\sim20$ before the epoch of re-ionization \cite{Volonteri10, Johnson16,Umeda16,Valiante16, 2017MNRAS.468.3935H, 2020ARA&A..58...27I, 2021MNRAS.500.4095V} and in the range of masses $\sim[10^4, 10^8] \msun$ in the nearby Universe. 
Detecting GWs from these sources will allow to reconstruct the merger history of MBHBs, disentangling the astrophysical processes and mechanisms driving their formation and evolution \cite{Sesana_reconstructing, Jaffe:2002rt, Wyithe:2002ep, Enoki:2004ew, Sesana:2004gf, Barausse12,PhysRevD.93.024003, 2020JCAP...11..055P,2022arXiv220306016A}, perform tests of general relativity \cite{PhysRevD.73.064030, Hughes:2004vw} and constrain cosmological scenarios \cite{Tamanini16,2017JCAP...05..031C,2016JCAP...10..006C,2022PhRvD.105f4061C,2021PhRvD.103h3526S,Belgacem19} (see \cite{2022arXiv220405434A,2022LRR....25....4A} for recent reviews on cosmological and fundamental physics implications of LISA).

Compact binaries emitting GWs can be considered ``standard sirens'', because they provide access to the source luminosity distance $d_L$. 
The latter is indeed encoded in the waveform and can be extracted directly, without resorting to a cosmic distance ladder, as necessary for type Ia supernovae (SNIa) \cite{doi:10.1146/annurev.astro.38.1.191}.
However, GWs alone do not provide the redshift of the source.
In the presence of an electromagnetic (EM) counterpart, the redshift can be obtained identifying and observing the host galaxy with EM facilities; this information can then be used to construct the $d_L-z$ diagram and constrain cosmological parameters \cite{Tamanini16, holz2005using} (see also \cite{Nissanke_2010,PhysRevD.86.043011, 2017Natur.551...85A} for ground-based detectors). If no EM counterpart is present, statistical methods can be employed to infer the cosmological parameters, if enough GW sources are available  \cite{schutz1986determining, 2011ApJ...732...82P,2018MNRAS.475.3485D,PhysRevD.95.083525, 2021arXiv210913934M,2021MNRAS.508.4512L,2022PhRvR...4a3247Z}.

In the case of MBHBs,
the presence of an EM counterpart accompanying the GW signal has long been discussed in the literature and the situation is still unclear, mostly due to the lack of observational evidence. In the presence of a sufficient amount of gas in the close environs of the binary, an EM counterpart can be triggered by the accretion of the gas onto the binary during the inspiral, merger or ringdown \cite{ROSA2020101525, Armitage_2002, Milosavljevic:2004cg, Dotti:2006zn, Kocsis_2006}. The binary motion is expected to excavate a cavity in the circumbinary disk, while gas streams from the inner edge of the disk should form minidisks surrounding each BH, contributing to spectral features and variable EM emission at various wavelengths previous to merger  \cite{10.1093/mnras/sty423,2014ApJ...785..115R, d_Ascoli_2018, 2015MNRAS.447L..80F, Franchini_2022}. Moreover, the orbital motion of the binary is expected to imprint a modulation on the EM counterpart from minidisks in phase with the GW signal, allowing for the  possible identification of the host galaxy in the field of view provided by LISA \cite{Bowen_2018, DalCanton19, McGee20}. Additional features can appear at or after merger, for instance an increase in jet power \cite{Palenzuela10_jet}, high accretion rate episodes similar to Active Galactic Nuclei (AGN) emission \cite{Milosavljevic:2004cg}, spectral or transient features caused by gravitational recoil \cite{2010MNRAS.401.2021R,2008ApJ...684..835S}.

In this work, we present different scenarios for the EM counterpart of MBHB mergers, exploring the potential of multimessenger observations with LISA and future EM facilities. 
This is the first of a series of  papers, sharing the common objective of upgrading  the analyses of \citet{PhysRevD.93.024003} and \citet{Tamanini16} (hereafter T16), to provide up to date forecasts on the ability of LISA to constrain MBHBs parameters (especially the sky position and the luminosity distance) with the final aim of probing the expansion of the Universe. For this reason, we present here detection strategies \emph{always including the redshift determination}.  Only in Sec.~\ref{subsec:mmcand_and_emcp_without_redshift}, for comparison, we provide the predicted numbers of MBHBs mergers with associated EM emissions without imposing the redshift determination.
Among the other papers, one will focus on the construction of the MBHBs standard sirens catalogues and on the inference of the cosmological parameters, while in the others we will discuss extensively the parameter estimation of the GW signal for this type of sources.

\section{General strategy}
\label{sec:general_strategy}

In order to provide updated forecasts for  multimessenger detection of MBHBs with LISA, we improve and complement the EM counterpart types proposed in T16, as well as the MBHBs parameter estimation with LISA.

Concerning EM counterparts, as put forward in T16, several options can be envisaged. 
First of all, if either the AGN or the host galaxy are sufficiently bright, and the LISA sky localization error small enough, the system can be identified and its redshift directly measured. 
Another possibility is the formation of a radio jet or flare during/after the merger, to be detected in the sky localization area provided by LISA. This would allow us to pinpoint the GW source sky position, with subsequent identification of the host galaxy. The source redshift could then be estimated either spectroscopically or photometrically with an optical telescope. Similarly, the X-ray emission associated to the MBHBs could also be used to identify the GW source sky position, and in turn to determine the host galaxy.


In the context of the counterpart types described above, we consider in this work specific EM observatories. For the direct  optical identification of an AGN at the time of the MBHB merger, we consider the Vera C. Rubin Observatory \cite{LSST, about_Rubin}.
We assume that the identification via the radio emission is performed by the future radio telescope Square Kilometre Array (SKA) \cite{ska}. In addition, we also explore the possibility to detect the X-ray EM counterpart with the Advanced Telescope for High ENergy Astrophysics (Athena) \cite{nandra13, Athena-synergies, chasing_mbhb}. Once the galaxy is identified from the radio or X-ray emission in the sky localization error region provided by LISA, the redshift measurement can be obtained with the Extremely Large Telescope (ELT) \cite{elt} as an example of a telescope with a $\gtrsim 30\rm \, m$ mirror or, if possible, directly with the Rubin Observatory.

In summary, we will analyse  3 observational scenarios:
\begin{enumerate}[label=(\alph*)]
    \item the Rubin Observatory alone (both identification and redshift)
    \item SKA (identification) + ELT (redshift)
    \item Athena (identification) + ELT (redshift)
\end{enumerate}
with several variations, detailed in Section \ref{sec:EM_counterpart} and \ref{sec:AGN-obscuration} to bracket the uncertainties. 

As a starting point, we need a population of merging MBHBs. Following T16, we adopt the result of semi-analytical models (SAM) \cite{Barausse12, Sesana14_spin_evolution,Antonini15_1, Antonini15_2} to track the evolution of BH masses, spins and surrounding gas across the cosmic time. Specifically we consider three models to explore different seed and time-delay prescriptions:
\begin{enumerate}
    \item  Pop3: a light-seed model with delays included where BHs form from very massive metal-poor stars at high redshift \cite{Madau01, Volonteri03};
    \item  Q3d: a heavy-seed model with delays included where MBHs originate from the collapse of protogalactic disks \cite{Volonteri08,Begelman06};
    \item  Q3nd: a similar heavy-seed model \cite{Volonteri08,Begelman06} but without delays between the galaxy and the BH merger, leading to more events but skewed toward higher redshift. In this sense, this scenario can be considered as optimistic in the number of predicted MBHB mergers. 
\end{enumerate}
These models predict the merger rate, the intrinsic binary properties (masses, spin magnitudes and orientations, luminosity distance) and the properties of the host galaxy (amount of mass in gas and stars, mass in the disk, etc.). For each model, we use a catalogue containing 90 years of data.
We assume 4 years of LISA observations, corresponding to an overall mission duration of 5 years with 80\% duty cycle of data taking. 
We further complete the catalogues by assigning randomly  to each event the sky position (uniform over the sky sphere), orbit inclination (random in $[0,\pi]$), polarization (random in $[0,2\pi]$), coalescence phase (random in $[0,2\pi]$), and merger time (random over $[0,1]$ years~\footnote{The orbits of LISA will not be perfectly periodic due to some fluctuations in the motion of the satellites and the degradation of the onboard instrumentation. However, for simplicity, we assume periodic orbits which allow us to rescale any arbitrary merger time in the interval $[0,1]$ yr. Moreover most of these systems will last less than 1 month in LISA band so the choice of the coalescence time does not impact significantly the parameter estimation. We also did not take into account signals truncated at the beginning and at the end of LISA mission}).

Given the simulated MBHB population, in order to reproduce the actual observational process, the next step should be to perform parameter estimation  of the GW signal for each of the MBHBs in the catalogues, to infer the sky localization error. 
If the latter is small enough, one would then turn to evaluating the detectability of the EM counterpart. 

In this work we use the Bayesian Markov Chain Monte-Carlo (MCMC) approach of \cite{Marsat21} for the LISA parameter estimation, improving on the Fisher forecast of T16.
However, this implies that the parameter estimation of the GW signal is the most computationally expensive step.
Moreover, not every system in the catalogues is expected to produce a detectable EM counterpart, as the emission might be too faint, or the merger might happen in a dry environment. 
Therefore, we choose to assess the  detectability of the EM counterpart in the first place. 
We select the systems whose fluxes (or magnitudes) are greater (smaller) than the corresponding threshold values for each of the listed EM facilities, and we run the parameter estimation only on this subset of events.
Applying an initial cut in the EM detectability allows us to reduce the number of sources for which we have to perform the parameter estimation, limiting the computational effort. 

For the subset of systems with detectable EM counterpart, we simulate the full inspiral-merger-ringdown GW signal in LISA using the waveform model PhenomHM for circularized binaries with aligned spins \cite{PhHM}, further select those with signal-to-noise ratio (SNR) above 10, and we estimate the binary parameters of these systems with the MCMC.

In order to detect an EM counterpart (especially if they are transients close to merger \cite{Armitage_2002}), telescopes must be pointed to the expected sky position of the GW source inferred by LISA (similarly to the alerts provided by LIGO/Virgo). For this reason, we apply a cut in the sky localization uncertainty. 
Together with the above mentioned SNR level and the detectability of the EM counterpart, we further require that the binary systems satisfy
 $\Delta \Omega < 10 \degsq$ to guarantee detection with the Rubin Observatory and SKA, or $\Delta \Omega < 0.4 \degsq$ to guarantee detection with Athena (see also Section~\ref{subsec:athena} for another possible strategy). 

Following this procedure, we define as \emph{multimessenger candidate} (hereafter MMcand in the figures) any MBHB system within the catalogue that satisfies the following two conditions:
\begin{enumerate}
    \item[] \emph{Multimessenger candidate}:
    \item The system has a detectable EM counterpart;
    \item The system has GW $ \rm SNR>10$.
\end{enumerate}
According to this definition, a multimessenger candidate is a system that can be detected by LISA and by any EM facility, but for which we impose no restrictive requirement on the sky localization. In other words, 
multimessenger candidates are mergers detectable by LISA for which the EM emission would be observable if the sky position were known with a certain accuracy.

We then define as \emph{GW event with EM counterpart} (hereafter EMcp) any system that satisfies the following two conditions:
\begin{enumerate}
    \item[] \emph{GW event with EM counterpart}:
    \item The system is a \emph{multimessenger candidate};
    \item The system is localized by LISA with $ \Delta \Omega < 10 \degsq$ if detectable by the Rubin Observatory and SKA, and/or $\Delta \Omega < 0.4 \degsq$ if detectable by Athena.
\end{enumerate}

We stress again that, according to the definition of the three observational strategies at the beginning of this section, we require the redshift determination both for multimessenger candidates and EMcps. 

An important caveat of our analysis is that the cuts in the number of GW events performed to implement the observability of the EM counterpart concern the magnitude level and the sky localization, but not the event sky position. 
This is particularly relevant for Earth-based observatories (i.e. the Rubin Observatory, SKA and ELT), which cover only a fraction of the sky.
The only way to implement a sky fraction cut would have been to apply an overall reduction factor on the number of GW events with EM counterpart, corresponding to the sky fraction  covered by each facility, and elaborate some technique to account for the detection of the same event by multiple telescopes. 
Instead of adopting this crude method, we have decided to neglect the observable sky fraction as a whole when we implement the detection with Earth-based telescopes.
This rather optimistic choice can lead to an overestimation of the number of predicted GW events with EM counterpart, with respect to those that might effectively be observable with actual data. We consider this a minor issue, compared with the uncertainty inherent in astrophysical MBHB evolution scenarios.

The paper is organised as follows. In Section~\ref{sec:MBHB_catalogue} we describe the MBHBs catalogues and the physics of the SAM that affects the formation and evolution of the MBHBs. 
In Section~\ref{sec:EM_counterpart} we present how we model the EM counterpart to MBHB mergers for different wavelengths and telescopes.
In Section~\ref{sec:AGN-obscuration} we describe a scenario where the EM flux is reduced by the surrounding gas. 
In Section~\ref{sec:GW_signal} we describe the tools we adopted to simulate the GW signal and to perform the parameter estimation. Our main results are reported in Section~\ref{sec:results}.
In Section~\ref{sec:multimodal_systems} we analyse multi-modal systems, i.e. systems whose sky localization inferred by LISA is dislocated in several portion of the sky. In Section~\ref{sec:discu_and_concl} we conclude with some final remarks and comments. In Appendix~\ref{sec:app_to_compare_with_Tamanini16} we compare our results with previous works in the literature, identifying the reasons of the discrepancies, when applicable. In Appendix~\ref{sec:app_parameter_estimation} we discuss briefly the role of the $\SNR$ and other binary parameters in determining the number of EMcps. Finally in Appendix~\ref{sec:app_plot_for_discussion} we present some figures useful for discussion.

\section{\label{sec:MBHB_catalogue}Catalogue of MBHBs}

The MBHB populations adopted in this work are based on a semi-analytical galaxy formation and evolution model \cite{Barausse12,Sesana14_spin_evolution, Antonini15_1, Antonini15_2} (the same model is employed also in \citet{Belgacem19} and in T16). We refer the interested readers to the original papers and here we summarise only the general features of the model. 

The model evolves dark matter merger trees from a Press-Schechter formalism along with the galactic baryonic structures, accounting for the complex interplay between the multiple components (intergalactic medium, interstellar medium, disk and nuclear cluster properties and the massive BH). 
Among the physical aspects that affect the mass distribution and the merger rate of MBHBs, we focus on two that have been shown to have a strong effect: the seed prescription, which defines the starting point for the MBH growth, and the time delay between the galaxy merger and the MBHB coalescence.

The light-seed prescription assumes that the first BHs form at high redshift ($z>15-20$) in the range $\sim[10^2, 10^3] \msun$ from the collapse of heavy Pop3 stars in the most metal-poor dark matter halos. We will refer to this seed model as `Pop3'. 
In the heavy-seed prescription, most of the mass in a protogalactic disk collapses into a supermassive star or a quasistar leaving behind a MBH  in the range $\sim[10^4, 10^5] \msun$ at $z\sim 8-15$. 
The heavy seeds are considered to be rarer than the light seeds due to their particular birth environmental conditions \cite{2017NatAs...1E..75R, 2016ApJ...832..134C}.

The time delays represent the time between the merger of the galaxies and the coalescence of the MBHBs. During a galaxy merger, the two MBHs migrate toward the center owing to dynamical friction \cite{Chandrasekhar1943} operating on the individual MBHs. Dynamical friction generated by the interaction between a massive perturber and the stellar and gaseous distribution decelerates the perturber. If dynamical friction is sufficiently effective, the MBH orbits will decay until they find each other at the center of the galaxy merger remnant and form a bound binary.  At this point the MBHs have typical separations of sub-parsec to a few parsecs, depending on the binary mass, still far from the orbital scale at which GWs can efficiently subtract energy and orbital angular momentum to the binary leading to the final coalescence ($\simeq 10^{-3} \rm \, pc$).
Additional processes are therefore needed to further shrink the orbit: energy exchanges in three-body interactions between the MBHB and nearby stars (a process referred to as stellar hardening), gas torques in circumbinary discs, and scattering between the MBHB and a third incoming MBH are included in the semi-analytical models of \cite{Antonini15_2}, which are used for this paper in order to account for how the MBHB crosses from parsec to milliparsec scales. 
There are large uncertainties in all these steps and assessing the efficiency  of these mechanisms is beyond the scope of this work. Here we summarise briefly how time delays have been derived in \cite{Antonini15_2} and we refer to T16 and \cite{Antonini15_2} for more details.
In a gas-rich environment with $M_{\rm res}> M_{\rm tot}$, where $M_{\rm res}$ is the mass of the reservoir gas available for accretion \footnote{ The radiation from young stars in the bulge is expected to bring the gas in a low-angular momentum disk surrounding the MBHB \cite{Haiman_2004} so, for our purposes, the accretion of the reservoir mass follow the star formation rate in the bulge.}  and  $M_{\rm tot} = m_1+m_2$ is the binary mass (in our work we set $m_1>m_2$ with $m_1$ the mass of the primary BH), the time delays are set by the viscous time of the nuclear gas (see Eq.~29 in \cite{Antonini15_2}), that is supposed to bring the MBHB to coalescence in $\lesssim 10^{7-8} \yr$. In the case of a gas-poor environment with  $M_{\rm res}< M_{\rm tot}$,  stars would be responsible for bringing the two MBHs together to the scale where GW emission is efficient (see Eq.~30-31 in \cite{Antonini15_2}). Furthermore in \cite{Antonini15_2}, the authors took also into account the role of three-body interactions that could lead to coalescence binaries that have stalled  (see Eq.~32 in \cite{Antonini15_2}). 

  The SAM tracks also the evolution of BHs spins. The spin magnitudes can increase (decrease on average) under coherent (chaotic) accretion. Moreover, after the merger, the spin magnitude and orientation of the remnant BH is computed according to the formalism described in \cite{2012ApJ...758...63B}. Since we model the GW signal with the PhenomHM  waveform \cite{PhHM} (see discussion in Sec.~\ref{sec:GW_signal}), we neglect the binary precession in the GW analysis.

%


\section{\label{sec:EM_counterpart}EM counterpart}
In this section we discuss the models of the EM counterparts. 
We upgrade and improve T16 in what concerns the optical emission of the AGN, which can
be detected with the Rubin Observatory, the radio jets, which can
be detected with SKA, and the near-IR galaxy
emission, which can be detected with ELT. 
To these counterpart types, we further add a model of the X-ray emission, to be detected by Athena.

The AGN bolometric luminosity $L_{\rm bol}$ is computed using the conventions in \cite{Merloni08}: 
\begin{align}
  L_{\rm bol} &= \min \left( \epsilon_{\rm rad}\dot{M}_{\rm acc} c^2,  L_{\rm Edd}\right) \label{eq:lbol}\\
  \dot{M}_{\rm acc} &= \min \left( \frac{M_{\rm res}}{{t_\nu}}, \frac{L_{\rm Edd}}{\epsilon_{\rm rad} c^2} \right)
  \label{eq:mdot}
\end{align} 
where $\epsilon_{\rm rad}$ is the radiative efficiency, $\dot{M}_{\rm acc}$ is the mass accretion rate and $L_{\rm Edd} = 4\pi G M_{\rm tot} m_p c /\sigma_T $ is the Eddington luminosity, $m_p$ the proton mass, $\sigma_T$ the Thompson cross section.

The SAM predicts the amount of gas surrounding the binary at merger. If the binary does not accrete at Eddington, $\dot{M}_{\rm acc} = M_{\rm res}/{t_\nu}$, where $t_\nu$ is the viscous timescale. The latter is computed as 
\begin{equation}
    t_{\nu} \simeq {\rm Re} \, t_{\rm dyn}
\end{equation}
where $\rm Re \approx 10^3$ is the critical Reynolds number and $t_{\rm dyn}$ is the dynamical timescale
\begin{equation}
    t_{\rm dyn} = G M_{\rm tot}/\sigma^3
\end{equation}
with $\sigma$ the velocity dispersion  
\begin{equation}
    \sigma = \sqrt{\frac{GM_{\rm b}}{3r_{\rm b}}} + \sqrt{\frac{GM_{\rm dyn}}{3 (2r_{\rm h})}}.
\end{equation}
In the above equation, $ M_{\rm b} $ is the total bulge mass from stars and gas, $ M_{\rm dyn} $ is the sum of the nuclear star cluster (NSC) mass and of the reservoir of gas feeding the MBH, $r_{\rm b}$ is the scale radius (assumed to be the same for gas and stars) and $r_{\rm h}$ is the NSC half light radii scale.
The scale radius $r_{\rm b}$ is related to the half-light radius $R_{\rm eff}$ as $r_{\rm b} = R_{\rm eff}/1.8153$ and $R_{\rm eff}$ is computed following Eq.~32 in \cite{Shen03}. The NSC half light radii scale $r_{\rm h}$ is computed instead as \cite{Harris96}
\begin{equation}
    r_{\rm h} = 3 {\rm pc} \,  \max\left[\sqrt{\frac{M_{\rm dyn}}{10^6 \msun}}, 1 \right].
\end{equation}

If the binary accretes at Eddington, on the other hand, in Eq.~\ref{eq:mdot} we limit the mass accretion rate to $\dot{M}_{\rm acc} =  L_{\rm Edd} / \epsilon_{\rm rad} c^2$. The radiative efficiency $\epsilon_{\rm rad}$ describes the fraction of rest frame energy that can be extracted by the matter accreting onto the MBH and depends on the accretion efficiency $\eta$ (the \emph{maximum} amount of energy that can be extracted) and on the accretion geometry. 
In other words, the radiative efficiency takes into account that not all the available energy is radiated but it can be used to produce jets or winds and its value might depend on structural changes in the accretion disc. 
The accretion efficiency $\eta$ depends on the spin of the MBH, $a_{bh}$, and ranges from 0.057 for Schwarzschild MBHs to $\sim0.4$ for maximally rotating MBHs.
If the merger is in a gas-rich environment with $M_{\rm res}> M_{\rm tot}$, we assume
\begin{equation}
    \eta(a) = 1- E_{\rm ISCO}(a),
\end{equation}
where $E_{\rm ISCO}$ is the specific energy around a rotating MBH with spin $a\in[-1,1]$ \cite{Bardeen1970}. 
If the binary is in a dry environment with $M_{\rm res}< M_{\rm tot}$, the accretion might happen in prograde and retrograde orbits with the same probability so the accretion efficiency becomes 
\begin{equation}
    \eta(a) = 1- \frac{1}{2}\bigg[E_{\rm ISCO}(a) + E_{\rm ISCO}(-a) \bigg].
\end{equation}
In our approach we choose $a=\chi_1$ where $\chi_1$ is the spin component of the primary MBH along the angular momentum.
The radiative efficiency is then computed as
\begin{equation}
    \epsilon_{\rm rad} = 
    \begin{cases}
    \eta(a)  &  \quad \dot{m} \ge \dot{m}_{\rm cr} \\
    \eta(a)\times \left( \frac{\dot{m}}{\dot{m}_{\rm cr}}\right)  &  \quad\dot{m} < \dot{m}_{\rm cr}
    \end{cases}
\end{equation}
being $\dot{m}_{\rm cr} = 0.01 $ and $\dot{m} = \eta M_{\rm res} c^2/ (L_{\rm Edd} {t_\nu})  $  \cite{Merloni08}. 

From the bolometric luminosity we can compute the absolute magnitude through \cite{Sparke-Gallagher}
\begin{equation}
\label{eq:abs_magn}
    M_{\rm band} = M_{\rm band, \odot} - 2.5\log_{10}\left( \frac{1}{BC} \frac{L_{\rm bol}}{L_{\odot}} \right)\,,
\end{equation}
where $M_{\rm band, \odot}$ is the absolute magnitude of the Sun in a certain band, $L_{\odot}$ the Sun luminosity and $BC$ the bolometric correction. From the absolute magnitude we can infer the apparent magnitude $m_{\rm band}$:
\begin{equation}
\label{eq:app_magn}
    m_{\rm band} = M_{\rm band} + 5\log_{10}\left( \frac{d_L}{\rm pc}\right)-5 +k_{\rm band}(z) \,,
\end{equation}
where the $k$-correction $k_{\rm band}(z)$ takes into account how the galaxy's radiation is redshifted during the propagation from the source to the observer.
In the optical band, we assume that the AGN spectrum is flat in $\nu f_{\nu}$ (see Fig.1 in \cite{bolometric_correction}), i.e $f_{\nu} \sim\nu^{-1}$. This approximation leads to $k_{\rm band}=0$ so we neglect the last term in Eq.~\ref{eq:app_magn}.


\subsection{The Vera C. Rubin Observatory}
\label{subsec:lsst}
The Rubin Observatory is an optical telescope with a $8.4 \rm \, m$ mirror for observations in the \emph{u}, \emph{g}, \emph{r}, \emph{i}, \emph{z}, \emph{y} bands and $9.6 \rm \, deg^2$ field of view (FOV). 

We envisage two possible counterpart detection strategies with the Rubin Observatory.
The Rubin Observatory Legacy Survey of Space and Time (LSST) is expected to reach a final depth of $m\sim 27.5$ after 10 years of operations in $r$ band \cite{LSST}. We therefore assume that we can identify the counterpart in the archival data, searching for a possible modulation due to the proper motion of the binary before merger. 
Moreover, LSST  will finish its current scientific objectives in $\sim$2032, so we do not exclude the possibility to carry a target of opportunity (ToO) observation for LISA candidates with the Rubin Observatory with an observation time of few hours. 
For simplicity we assume the same apparent magnitude threshold for the ToO as for the entire survey. We also use the subscript `Rubin' when the equations refer both to the LSST or ToO case.

For the AGN spectral energy distribution (SED) in optical bands we assume a bolometric correction $BC =10$   (Fig.~2 in \cite{bolometric_correction} shows that the bolometric correction is almost constant around 10 in optical bands) and compute the absolute magnitude following Eq. \ref{eq:abs_magn} as 
\begin{equation}
    M_{\rm AGN,  Rubin} = 4.64 - 2.5\log_{10}\left(  \frac{0.1 L_{\rm bol}}{L_{\odot}} \right)\,,
\end{equation}
where we have inserted $M_{\rm band, \odot} = 4.64$, the Sun AB magnitude in $r$ band \cite{Solar_magnitude}.
The apparent magnitude $m$ becomes then
\begin{equation}
\label{eq:magn_lsst}
    m_{\rm AGN,  Rubin} = M_{\rm AGN,  Rubin} + 5\log_{10}\left( \frac{d_L}{\rm pc}\right)-5.
\end{equation}
We fix the threshold magnitude for detection with LSST and with ToO with the Rubin Observatory at $m_{\rm AGN, Rubin, \, lim} = 27.5$ and claim detection of the multimessenger candidate if
\begin{equation}
    m_{\rm AGN,  Rubin} < m_{\rm AGN, Rubin, \, lim}.
\end{equation}
Once the source is identified, we can get an accurate redshift determination via photometric measurements, with error on the redshift $\Delta z = 0.031(1+z)$ \cite{Laigle19}. Concerning the sky localization threshold, we adopt the value of $\Delta \Omega = 10 \degsq$, close to the Rubin Observatory FOV.

\subsection{The Square Kilometre Array}
\label{subsec:ska}

When LISA will be taking data, the SKA will be the largest radio telescope on Earth, with more than a square kilometer of collecting area. 
We therefore include the possibility of ToO with SKA in the proposed scenarios for EM counterpart detection.

During the MBHB merger, the interaction between the surrounding plasma and the magnetic fields is expected to produce radio emission \citep{Palenzuela10_jet,Gold14,Kelly17,2021PhRvD.103j3022C}. In particular, the motion of the binary is expected to twist the magnetic field lines, leading to flare emissions, while the Blandford-Znajeck effect \cite{Blandford-Znajek} could be responsible for strong radio jets, depending on the amount of accreted material.

Following T16, we model the flare emission as
\begin{equation}
\label{eq:Lflare}
    L_{\rm flare} = \frac{ \epsilon_{\rm Edd} \epsilon_{\rm radio}}{ q^2} L_{\rm Edd}
\end{equation}
where $\epsilon_{\rm Edd} = L_{\rm bol}/L_{\rm Edd} $ is the Eddington ratio, $q = m_1/m_2 $ is the binary mass ratio and $\epsilon_{\rm radio} = 0.1$ is the portion of bolometric EM radiation emitted in radio band.
In our model, $\epsilon_{\rm Edd}$ is either computed from the reservoir amount of gas surrounding the binary at merger, with a floor value of $\epsilon_{\rm Edd}= 0.02$, or it is set equal to 1 if the binary accretes at Eddington.

Furthermore, the jet luminosity is modelled as \cite{Barausse12, Meirer01}
\begin{widetext}
\begin{equation}
\label{eq:Ljet}
    L_{\rm jet} = \begin{cases} 0.8 \times  10^{42.7}\, {\rm erg \, s^{-1}} m_9^{0.9} \left( \frac{\dot{m}_{\rm jet}}{0.1}\right)^{6/5} \left( 1+1.1a_1+0.29a_1^2  \right), & \mbox{if }10^{-2} \le \epsilon_{\rm Edd} \le 0.3\\ 3 \times 10^{45.1}\, {\rm erg \, s^{-1}} m_9 \left( \frac{\dot{m}_{\rm jet}}{0.1}\right)g^2 \left( 0.55f^2+1.5fa_1+a_1^2  \right) & \mbox{otherwise}
\end{cases} 
\end{equation}
\end{widetext}
where $m_9 = m_1/(10^9 \msun)$,  $ \dot{m}_{\rm jet} =  \dot{M}_{\rm acc}/(22\, m_9 \msun \yr^{-1}) $ is the accretion rate in Eddington limit unit, $a_1$ is the spin magnitude of the primary black hole, $f=1$ and $g=2.3$ are dimensionless parameters regulating the angular velocity and azimuthal magnetic field of the system. The jet luminosity is the only emission that could reach $\sim 10 L_{\rm Edd}$. $L_{\rm flare}$, as well as the other emissions considered in this work (in optical and X-ray) are Eddington limited by construction.

Radio jets are expected to be beamed with an opening angle $\theta  = 1/\Gamma$,  $\Gamma $ being the Lorentz factor \cite{Blandford79, Yuan20}. 
Two competing effects come into play: 
\begin{itemize}
    \item If the line of sight is outside the cone angle of the jet, the radio emission cannot be detected;
    \item Collimated emission increases the flux received from the observer, allowing for the detection of fainter and farther systems.
\end{itemize}
For typical AGNs, $\Gamma \simeq 5-15$ leading to a corresponding opening angle of $\theta \simeq 6^\circ $. 
 \citealt{Yuan20}, however, assumed instead a fiducial value of $\Gamma =2$ based on the simulations of \cite{Gold14}, corresponding to $\theta \simeq 30^\circ $. 
If the emission is beamed, the luminosity increases as \cite{Cohen07}
\begin{eqnarray}
L_{\rm beamed}= L_{\rm isotropic} \delta^n\,,\\
\delta = \Gamma^{-1} \left(1-\beta \cos \iota  \right)^{-1}\,,\\
\beta = \sqrt{1 - \frac{1}{\Gamma^2}}\,, 
\end{eqnarray}
where $\beta$ is the beam speed in the AGN frame in units of the speed of light, $\iota$ is the inclination angle between the binary angular momentum and the line of sight and $n=2$ is an index describing the geometry and spectral index of the jet. For simplicity, we assume  that the jet is aligned with the binary angular momentum.
We consider three scenarios for the radio counterparts:
\begin{itemize} 
    \item `Isotropic flare': the flare emission is isotropic and the jet is collimated with $\Gamma=2$. In this case we compute the total luminosity in the radio band as
    \begin{equation}
    L_{\rm radio} = \begin{cases} 
    L_{\rm flare}+L_{\rm jet}^{\rm beamed} & \mbox{if } \iota <30^\circ \mbox{ or } \iota>150^\circ \\ 
    L_{\rm flare} & \mbox{otherwise;}
    \end{cases}
    \end{equation}
    
    \item `$\Gamma 2$' : both the jet and the flare are beamed with $\Gamma = 2$. The total luminosity is the sum of the flare and jet collimated emissions, or it is zero if we are outside the cone:
    \begin{equation}
    L_{\rm radio} = \begin{cases} 
    L_{\rm flare}^{\rm beamed}+L_{\rm jet}^{\rm beamed} & \mbox{if } \iota <30^\circ \mbox{ or } \iota>150^\circ \\
    0 & \mbox{otherwise;}
    \end{cases}
    \end{equation}
    
    \item `$\Gamma 10$' : the same as `$\Gamma = 2$' but setting $\Gamma = 10$. The total luminosity is computed as:
    \begin{equation}
    L_{\rm radio} = \begin{cases} 
    L_{\rm flare}^{\rm beamed}+L_{\rm jet}^{\rm beamed} & \mbox{if } \iota <6^\circ \mbox{ or } \iota>174^\circ \\
    0 & \mbox{otherwise.}
    \end{cases}
    \end{equation}
    
    
\end{itemize}

For each of the three scenarios above, we use $L_{\rm radio}$ to calculate the flux
\begin{equation}
    F_{\rm radio} =\left( \frac{L_{\rm radio}}{\rm erg/s}\right) \left(\frac{d_L}{\rm cm} \right)^{-2}.
\end{equation}
To detect the multimessenger candidate with SKA, we impose that
\begin{equation}
\label{eq:detection_ska}
   F_{\rm radio} \ge 4 \pi 10^{-20} F_{\rm lim}^{\rm SKA},
\end{equation}
where  $F_{\rm lim}^{\rm SKA} = \nu_{\rm SKA}  F_{\rm \nu, \, lim}^{\rm SKA} $ is the minimum flux detectable with SKA, $\nu_{\rm SKA} = 1.7 \rm \, GHz $ is the typical frequency at which we expect the bulk emission and $F_{\rm \nu, \, lim}^{\rm SKA} = 1 \mu \rm Jy$ is the flux limit in the same band. The final configuration of SKA is expected to reach this sensitivity in a FOV of $\sim 10 \degsq$ and $\sim 10$ minutes of integration time. We note that this is an optimistic approach that assumes that all the Poynting flux is converted into jet radio emission, which is not guaranteed \cite{2012ApJ...749L..32M}.

Note that the radio jet luminosity can be also computed from the fundamental plane relation \cite{Gultekin19}. Ongoing work is evaluating how the fundamental plane relation compares with Eq.~\ref{eq:Ljet} \cite{Chi_An_paper}.
Here we have verified that both methods to calculate $L_{\rm jet}$ lead to a similar number of radio EM counterparts, since the flare luminosity on its own is generally detectable with an SKA flux limit of 1 $\mu \rm Jy$ .

For SKA's sky localization threshold we adopt the same limit as in T16, i.e. $\Delta \Omega = 10 \degsq$. 

\subsection{Athena}
\label{subsec:athena}

Being the 2nd Large class mission of the European Space Agency, the X-ray telescope Athena will observe the most energetic events from the high-redshift Universe thanks to its Wide Field Imager (WFI) with a FOV of $0.4 \degsq$ \footnote{At the moment of writing this paper, there have been rumors concerning the possibility of Athena mission being descoped. Currently nothing is yet decided so we chose to adopt the current Athena configuration but we plan to review our result if any major issue will rise. We also discuss the value of having LISA and Athena operational at the same in Sec.~\ref{sec:discu_and_concl}}.

Together with the radio and optical/IR emission, MBHs produce copious amount of X-ray radiation. 
In this work we consider a late-stage X-ray emission produced by the gas accretion into the newly formed remnant MBH, possibly leading to a re-brightening of the source \cite{Milosavljevic:2004cg}. 
The increase in X-ray flux would allow the identification of the binary, among the possible X-ray transient candidates in the LISA sky localization area. 
In addition, internal shocks between the inner disk, closer to the remnant BH, and the outer disk portion are expected to produce X-ray emissions, even if they might be too faint to be detectable \cite{2010MNRAS.401.2021R}. 

We focus on X-ray emission in the soft band, between $0.5-2$ keV, because Athena is most sensitive in this band.
The bolometric correction in the X-ray band is given by \cite{bolometric_correction}
\begin{equation}
\label{eq:bol_correction}
    L_{\rm X} = \frac{L_{\rm bol}}{c_1 \left( \frac{L_{\rm bol}}{10^{10} L_{\odot}} \right) ^{k_1} + c_2 \left( \frac{L_{\rm bol}}{10^{10} L_{\odot}} \right) ^{k_2} }
\end{equation}
where $L_{\rm bol}$ is defined in Eq.~\eqref{eq:lbol}, $c_1 = 5.712$, $k_1 = -0.026$, $c_2 = 17.67$ and $k_2 = 0.278$.
For the MBH accretion, we consider two possibilities: 
\begin{itemize}
    \item The MBH accretion remains at the same level as before the merger, and we estimate it with the left term in Eq.~\ref{eq:mdot};
    \item The sudden disappearance of the torques from the binary leads to the infall of the gas in the disc,  and accretion reaches the Eddington luminosity.
\end{itemize}
The X-ray flux is 
\begin{equation}
\label{eq:flux_xray}
    F_{\rm X} = \left( \frac{L_{\rm X}} {\rm erg \, s^{-1}}\right)
    \frac{1}{4 \pi (d_L/ {\rm cm})^2}
\end{equation}
where $L_{\rm X}$ is the X-ray luminosity calculated with Eq.~\ref{eq:bol_correction}.

We define two possible detection strategies with Athena.
As the exposure time increases, Athena will be able to detect fainter sources. 
We set a reasonable maximum exposure time of $300 \rm \, ks$ \cite{guainazzi_personal} and two different flux limits, chosen together with appropriate LISA sky localization cuts:
\begin{itemize}
    \item If Athena observes the same FOV of $\Delta \Omega = 0.4 \rm \, deg^2$ for the entire exposure time, we assume a limiting flux of $F_{\rm X, \, lim} = 4 \times 10^{-17} \rm erg \, s^{-1} \, cm^{-2} $;
    \item Athena could also explore a larger sky localization, limited however to brighter flux. For this second strategy, we choose a flux limit of $F_{\rm X, \, lim} = 2 \times 10^{-16} \rm erg \, s^{-1} \, cm^{-2} $ and a sky localization threshold of $\Delta \Omega = 2 \rm \, deg^2$. This corresponds to $\sim 5$ tiles in the sky, each covered twice.
\end{itemize}

We claim detection of the X-ray counterpart whenever
\begin{equation}
    F_{\rm X} \ge F_{\rm X, \, lim}.
\end{equation}

In the second detection strategy, we do not take into account the slew time necessary to repoint Athena, because the tiles are next to each other (see Section~\ref{sec:multimodal_systems} for more details on LISA sky localization).
Finally, for simplicity, in this work we do not consider the possibility of pre-merger modulated X-ray emission \cite{DalCanton19}.


\subsection{The Extremely Large Telescope}
\label{subsec:elt}

Since one of our aims is to use LISA EMcps to test the expansion of the Universe, we also enforce the redshift measurement of the GW source.
Radio and X-ray observations will identify the sky localization of the merger event within the LISA sky area, while the redshift information relies on emission lines at optical/IR wavelengths. 
The simplest option is to detect the emission of the galaxy hosting the MBHB merger.

By the time LISA will be operative, the Extremely Large Telescope (ELT) will be available and other similarly large telescopes, such as the Giant Magellan Telescope and the Thirty Meter Telescope, are also under discussion. We focus the discussion here on ELT as a case study.  Among the ELT instruments, the spectrograph MICADO \cite{MICADO} will allow redshift measurements in the IYJHK band with spectroscopy between the OH lines down to an apparent magnitude of 27.2  and with imaging with advanced filters down to an apparent magnitude of 31.3. These values correspond to the 5$\sigma$ sensitivity for isolated point sources in 5 hours of observation.

We compute the galaxy luminosity in K band as
\begin{equation}
    L_{\rm gal, ELT} = \frac{1}{\Upsilon}L_{\odot} M_{\rm stars}\,,
\end{equation}
where $M_{\rm stars}$ is the total mass of the stars in the disk and bulge, and $\Upsilon$ is a fiducial mass-to-light ratio.
The latter requires information, such as the star formation history and the metallicity, which is not provided by the MBHB catalogues. 
We therefore adopt a simple correction. 
For systems at $z\ge3$ we assume $\Upsilon = 0.03$, because galaxies are dominated by young stellar populations with low metallicity, so that radiation in the $H$, $J$ and $K$ bands in the observer's rest frame actually comes from the blue part of the source's restframe spectral energy distribution. 
For systems with $z<3$, we assume $\Upsilon = 0.1$ because the stars are older and more metallic, and the observer-frame wavelength comes from the optical part of the restframe spectrum. 
Following Eqs.~\ref{eq:abs_magn}-\ref{eq:app_magn} the apparent magnitude is 
\begin{align}
  M_{\rm gal, ELT} &= 5.08 - 2.5\log_{10}\left( \frac{L_{\rm gal, ELT}}{L_{\odot}}\right) \label{eq:abs_magn_elt}\\
  m_{\rm gal,ELT} &= M_{\rm gal,ELT} + 5\log_{10}\left( \frac{d_L}{\rm pc}\right)-5.
  \label{eq:app_magn_elt}
\end{align}

For systems with $m_{\rm gal, ELT}\leqslant 27.2$  we expect that a spectroscopic redshift measurement will be doable, with precision $\Delta z = 10^{-3}$. This relies on the reasonable assumption that  galaxies are star-forming and therefore emission lines are present, since
the typical galaxies in the catalogue have masses of $10^7-10^{11} \msun$ at $z>1$. 

For galaxies with apparent luminosity $27.2<m_{\rm gal, ELT}<31.3$, the redshift measurement and its uncertainty  are less straightforward, and depend on the actual galaxy redshift. 
For high-redshift sources at $z\gtrsim6.5$, MICADO will be able to detect the Lyman-$\alpha$ (1215.67 \AA) break in the I-band, enabling the redshift measurement with error $\Delta z = 0.2$ \cite{Dunlop12}. 
For galaxies between $5<z\lesssim6.5$, the redshift measurement is more challenging, also because they might resemble ultra-faint galaxies at $z<0.5$. 
First of all, we assume that this degeneracy can be broken by using the information on the luminosity distance inferred from the GW measurement, in the context of a given cosmology. 
Even though the aim is to use the redshift information to \emph{infer} the cosmology, we believe that assuming the cosmology with the only aim of discriminating between sources at $z<0.5$ and $z>5$ will not substantially bias the cosmological analysis.
Furthermore, the Roman Space Telescope \cite{Roman} will  observe also in the R band, corresponding to the Lyman-$\alpha$ break for sources at $z\gtrsim5.2$. 
There will be the possibility to perform ToO follow-up within 2 weeks from notification, but with a limiting magnitude of 28.5 in 1hr (but longer observing time might be possible) \footnote{We report this value for completeness and we do not use this value in the analysis. Therefore we do not require the detectability also with the Roman Space Telescope. More details can be found at \url{https://roman.gsfc.nasa.gov/instruments_and_capabilities.html}}. 
Consequently, assuming observations with the Roman Telescope combined with the luminosity distance information from GW sources, we claim that the redshift identification will also be possible for sources with $z>5$ with uncertainty $\Delta z=0.2$, thanks to the detection of the Lyman-$\alpha$ break. 

Finally, for systems with $27.2<m_{\rm gal, ELT}<31.3$ and $0.5<z<5$, the photometric redshift can be obtained thanks to the Balmer break, 
with the penalty, however, of observing  only in near-IR bands (from I to K with MICADO): 
missing the optical and UV parts of the spectrum will increase the  photometric redshift errors. 
Unfortunately, we did not find reliable estimates of the redshift measurement error in the literature for this kind of sources with only near-IR observations. 
Therefore, we adopt an agnostic estimate and set $\Delta z = 0.5$ for these systems \footnote{In paper II we will provide more detail on this point and show that our cosmological results are robust against this, somewhat arbitrary, choice of the redshift error for systems with $27.2<m_{\rm ELT}<31.3$ and $0.5<z<5$.}.

Tab.~\ref{tab:elt_strategy} summarizes the errors adopted for redshifts measured based on the host galaxy spectra using ELT and the Roman Space Telescope.
Note that we do not take into account the possibility of observations with the James Webb Telescope \cite{James-Webb}, because it may not overlap with LISA.

\begin{table}
\caption{\label{tab:elt_strategy} Summary of the redshift errors adopted for ELT }
\begin{ruledtabular}
\begin{tabular}{c | m{2.4cm} | m{3.4cm}  } 
   & $m_{\rm gal,ELT}<27.2$  & $27.2<m_{\rm gal,ELT}<31.3$  \\
\Xhline{0.1pt}
$z \leqslant 0.5$   &   & No redshift information  \\
$0.5<z \leqslant 5$  & $\Delta z = 10^{-3}$ & $\Delta z = 0.5$  \\
$z>5$  &  & $\Delta z = 0.2$  \\

\end{tabular}
\end{ruledtabular}

\end{table}

\section{\label{sec:AGN-obscuration}AGN obscuration}

The gas and dust surrounding the MBHs are expected to absorb a fraction of the EM emission, reducing the observed flux and the number of detected systems \cite{Almeida17,Buchner16}.  Similarly, interstellar dust and intergalactic gas can absorb starlight in the galaxy.  

In this section we present how we model obscuration at X-ray and optical wavelengths, affecting the detection of the EM emission with Athena and with the Rubin Observatory respectively. We also model absorption from the galaxy gas, affecting the optical emission detectable with ELT (used to determine the EMcp redshift). 
Below we detail the two in turns, starting with the AGN obscuration.

Evaluating the fraction of obscured AGNs with respect to the total number of AGNs is generally challenging, and there is no consensus on how it evolves as function of luminosity or redshift \cite{2007A&A...463...79G,2014MNRAS.437.3550M,2015ApJ...802...89B}. Nevertheless, we adopt here a recent modelling of obscuration in X-rays, taken from \cite{Ueda14}. 
We further include absorption in the optical, assuming that dust follows gas.

For each event, we start by computing the hydrogen column density around the MBH from equations 3-6 in \cite{Ueda14}. For completeness, we detail here the entire procedure.
We introduce the $\psi(L_{X, h}, z)$ parameter that corresponds to the fraction of absorbed Compton-thin AGNs with respect to the total Compton-thin AGNs:
\begin{equation}
\begin{split}
  \lefteqn{ \psi(L_{X, h}, z) = \min[ \psi_{\rm max}, \max[\psi_{43.75}(z)} & \\ 
 & - \beta (\log_{10}[L_{X, h}/({\rm erg/sec})] - 43.75), \psi_{\rm min}] ]\,,
\end{split}
\end{equation}
where $\psi_{\rm max} = 0.84$, $\psi_{\rm min} = 0.2$, $\beta = 0.24$. 
The source redshift $z$ and the X-ray luminosity in hard band $L_{X, h}$ are the only input parameters, and we infer them from the information contained in the MBHBs catalogues. 
To compute $L_{X, h}$ we use Eq.~\ref{eq:bol_correction}, with coefficients
$c_1 = 4.073$, $k_1 = -0.026$, $c_2 = 12.60$ and $k_2 = 0.278$ \cite{bolometric_correction}. 
The quantity $\psi_{43.75}(z)$ represents the fraction of Compton-thin AGNs with $\log_{10} [ L_{X, h}/{\rm (erg/s)} ]= 43.75$ at $z$, and it takes the form
\begin{equation}
    \psi_{43.75}(z) = \begin{cases}
     0.43\,(1+z)^{0.48} & z < 2\,, \\
     0.43\,(1+2)^{0.48} & z \ge 2\,. \\
    \end{cases}
\end{equation}
The distribution of the hydrogen column density $N_H$ can be expressed as 
\begin{eqnarray}
   \lefteqn{f(L_{X, h}, z; N_H) =} & \\ 
   & =\begin{cases}
    1-\frac{2+\epsilon}{1+\epsilon}\psi(L_{X, h}, z) & 20 \le \log_{10} \left(\frac{N_H}{{\rm cm^2}} \right) <21 \\
    \frac{1}{1+\epsilon}\psi(L_{X, h}, z) & 21 \le \log_{10} \left(\frac{N_H}{{\rm cm^2}} \right) <23 \\
    \frac{\epsilon}{1+\epsilon}\psi(L_{X, h}, z) & 23 \le \log_{10} \left(\frac{N_H}{{\rm cm^2}} \right) <24 \\
    \frac{f_{\rm AGN}}{2}\psi(L_{X, h}, z) & 24 \le \log_{10} \left(\frac{N_H}{{\rm cm^2}} \right) <26 \nonumber
    \end{cases} \\
      &    {\rm if}~~\psi(L_{X, h}, z) < \frac{1+\epsilon}{3+\epsilon}\,,\nonumber
\end{eqnarray}
 and
\begin{eqnarray}
    \lefteqn{f(L_{X, h}, z; N_H) =} & \\ & =\begin{cases}
    \frac{2}{3}-\frac{3+2\epsilon}{3+3\epsilon}\psi(L_{X, h}, z) & 20 \le \log_{10} \left(\frac{N_H}{{\rm cm^2}} \right) <21 \\
    \frac{1}{3} - \frac{\epsilon}{3+3\epsilon}\psi(L_{X, h}, z) & 21 \le \log_{10} \left(\frac{N_H}{{\rm cm^2}} \right) <22 \\
    \frac{1}{1+\epsilon}\psi(L_{X, h}, z) & 22 \le \log_{10} \left(\frac{N_H}{{\rm cm^2}} \right) <23 \\
    \frac{\epsilon}{1+\epsilon}\psi(L_{X, h}, z) & 23 \le \log_{10} \left(\frac{N_H}{{\rm cm^2}} \right) <24 \\
    \frac{f_{\rm AGN}}{2}\psi(L_{X, h}, z) & 24 \le \log_{10} \left(\frac{N_H}{{\rm cm^2}} \right) <26 \nonumber
    \end{cases}\\
    &    {\rm if}~~\psi(L_{X, h}, z) \ge \frac{1+\epsilon}{3+\epsilon}\,,\nonumber
\end{eqnarray}
where $f_{\rm AGN} = 1 $ is the fraction of Compton-thick AGNs to the absorbed Compton-thin AGNs, and $\epsilon = 1.7$. Given the hard X-ray luminosity and redshift of each binary in the catalogues, we derive, using the equations above, the corresponding hydrogen column density distribution $f$. 
We then sample this distribution, in order to extract a value for the hydrogen column density $N_H$ surrounding the binary.

We then need to evaluate the absorption. 
We consider the soft X-ray band and compute the luminosity after absorption as
\begin{equation}
    L_{\rm X, abs} = L_{\rm X} \, e^{-\tau_{\rm X}}
\end{equation}
with $\tau_{\rm X} = \sigma_{\rm X} N_H$, where $\sigma_{\rm X}$ is the X-ray cross section, for which we take, following \cite{1983ApJ...270..119M},
\begin{equation}
\begin{split}
\sigma_{\rm X}   = & \left( 120.6 + 169.3(E/{\rm keV})  -47.7(E/{\rm keV})^2 \right) \\
&(E/{\rm keV})^{-3} \times 10^{-24} {\rm cm^{2} }.
\end{split}
\end{equation}
For the energy, we choose $E = 1 \, \rm keV$, in the middle of the soft X-ray band. The X-ray flux after obscuration can be computed from Eq.~\ref{eq:flux_xray}, substituting the original $L_{\rm X}$ with $L_{\rm X, abs}$.  We are applying a statistical correction based on observational samples of AGNs. An alternative approach could be to model obscuration based on the intrinsic properties of the source. For instance \citealt{2017Natur.549..488R} found in a low-redshift AGN sample a relation between the obscuration fraction and the Eddington ratio that can be used to compute the corresponding hydrogen column density, under the assumption that radiation pressure is the dominant factor modulating obscuration, caused by material very close to the black hole. In high redshift sources, however, the interstellar medium can also contribute to obscuration \cite{2019MNRAS.487..819T,2019A&A...623A.172C}. We prefer therefore to base our correction on empirical results based on observational samples covering a broad range of redshifts. 

Similarly as for X-ray, the AGN emission in the optical band after absorption is 
\begin{equation}
\label{eq:lum_lsst_abs}
    L_{\rm bol, abs} = L_{\rm bol} \, e^{-\tau_{\rm opt}}\,,
\end{equation}
where $L_{\rm bol}$ is the bolometric luminosity from Eq.~\ref{eq:lbol}, and $\tau_{\rm opt}=\sigma_{\rm opt}N_{\rm dust}$ is the optical depth, with $\sigma_{\rm opt}$ the optical cross section and $N_{\rm dust}$ the dust column density.
In order to evaluate the optical depth, we start from the galaxy mass-metallicity relation $Z_{\rm gas}$ \cite{2016MNRAS.456.2140M}
\begin{equation}
\begin{split}
    \log_{10}( Z_{\rm gas}/ Z_{\odot}) &= 0.35 [ \log_{10}(M_{\rm stars}/ M_{\odot}) -10   ] \\ & + 0.93 {\rm e}^{-0.43z} - 1.05.
\end{split}
\end{equation}
From the hydrogen column density, calculated as described above, and $Z_{\rm gas}$, one obtains the dust column density \cite{Gnedin08}:
\begin{equation}
\label{eq:ndust}
    N_{\rm dust} = \frac{Z_{\rm gas}}{0.02}N_H 
\end{equation}
and the optical cross section, given by 
\begin{equation}
\label{eq:optical_sigma}
    \sigma_{\rm opt} = \sigma_0 \sum_{i=1}^7 F(\lambda/\lambda_i, a_i, b_i, p_i, q_i)\,,
\end{equation}
where $\sigma_0 = 3 \times 10^{-22} \, \rm{cm^2}$ and the fitting function $F$ as
\begin{equation}
    F(x, a, b, p, q) = \frac{a}{x^p + x^{-q} +b}\,,
\end{equation}
with coefficients $\lambda_i, a_i, b_i, p_i, q_i$ reported in Tab.~\ref{tab:coeff_sigma_optical}. 
We choose to compute the cross section at a reference wavelength of $\lambda = 0.62 \, \mu \rm{m}$, at the center of \emph{r} band.
The magnitude after absorption can be inferred from Eq.~\ref{eq:abs_magn}, substituting  $L_{\rm bol}$ with $L_{\rm bol, abs}$.

At last, we turn to the absorption of the optical galactic emission from interstellar dust and intergalactic gas, to be accounted for in the detection of the host galaxy by ELT. 
Based on  Fig.~1 in \cite{Tanvir19}, we adopt a constant hydrogen column density $\log_{10}(N_H/\rm cm^2) = 22 $ and compute the absorbed luminosity in K band following Eqs.~\ref{eq:lum_lsst_abs}-\ref{eq:ndust}, with $\lambda = 2.2 \, \mu \rm{m}$, at the center of K band. 
The absorption of the host galaxies emission does not affect significantly the EMcps rates, and it is always included in the following results.

\begin{table}
\caption{\label{tab:coeff_sigma_optical} Coefficients entering the optical cross section in Eq.~\ref{eq:optical_sigma} }
\begin{ruledtabular}
\begin{tabular}{m{1cm}|m{1.3cm}|m{0.9cm}|m{0.9cm}|m{0.6cm}|m{0.6cm} } 
\centering Term   & \centering $\lambda_i$ [$\mu$m]  & \centering $a_i$  & \centering $b_i$ & \centering $p_i$ &  $q_i$ \\
\Xhline{0.1pt}
\centering 1  & 0.042  & 185 & 90 & 2 & 2 \\
\centering 2 & 0.08 & 27 & 15.5 & 4 & 4 \\
\centering 3 & 0.22 & 0.005 & -1.95 & 2 & 2 \\
\centering 4 & 9.7 & 0.01 & -1.95 & 2 & 2 \\
\centering 5 & 18 & 0.012 & -1.8 & 2 & 2 \\
\centering 6 & 25 & 0.03 & 0 & 2 & 2 \\
\centering 7 & 0.067 & 10 & 1.9 & 4 & 15 \\
\end{tabular}
\end{ruledtabular}
\end{table}

\section{\label{sec:GW_signal}Gravitational wave signal and parameter estimation}

The waveform of a MBHB with aligned spins  and circular orbit depends on 11 parameters: the primary and secondary source-frame masses, $m_1$ and $m_2$, the two spins magnitudes along the orbital angular momentum, $\chi_1$ and $\chi_2$, the sky latitude $\beta$ and the longitude $\lambda$, the luminosity distance $d_L$, the inclination $\iota$ of the binary angular momentum with respect to the line of sight, the phase at coalescence (or at a reference frequency) $\phi$ , the time at coalescence $t_c$, and the polarization angle $\psi$.
We can also define the mass-ratio $q = m_1/m_2 \geqslant 1$ and the chirp mass $\mathcal{M} = (m_1 m_2)^{3/5}/(m_1 +m_2)^{1/5}$.
We model the GW signal with the  inspiral-merger-ringdown waveform PhenomHM \cite{PhHM} that ignores the binary spins precession but includes higher order harmonics.

If the source is located at cosmological distance, the signal is affected by the expansion of the Universe. 
As a consequence, in the detector-frame we measure redshifted quantities, i.e. $\mchirp_z = \mchirp (1+z)$, where $z$ is the source redshift. 
However, in this work we will refer to rest-frame quantities, unless otherwise stated. 
We assume a fiducial $\Lambda$CDM cosmology with $h = 0.6774$, $\Omega_m = 0.3075$ and $\Omega_{\Lambda} = 0.6925$.

For each system we compute the signal-to-noise ratio as 
\begin{equation}
    {\rm SNR} = 4 \int_{f_{\rm min}}^{f_{\rm max}} \frac{|\tilde{h}(f)|^2 }{S_n(f)} df\,,
\end{equation}
where $\tilde{h}(f)$ corresponds to the Fourier transform of the time-domain signal, and $S_n(f)$ is the noise power spectral density, for which we take the estimate ``SciRDv1'' described in \cite{Stas_sensitivity}. 
We set $f_{\rm min} = 10^{-5} \rm \, Hz$ and $f_{\rm max} = 0.5 \rm \, Hz$.
We assume 5 years of overall mission duration, with 80\% duty cycle. 
We add to the LISA noise PSD the one of the confusion background from unresolved galactic binaries, according to the fits presented in \cite{2021PhRvD.104d3019K}. The amplitude of the background is taken for three years of mission duration, as a representative average value over the total duration of 5 years. 

For each event we obtain the posterior distribution $p(\overline{\theta}|d)$ for a set of parameters $\overline{\theta}$  following Bayes theorem, 
\begin{equation}
p(\overline{\theta}|d) = \frac{\mathcal{L}(d|\overline{\theta}) \pi(\overline{\theta})}{p(d)}\,,
\end{equation}
where $\mathcal{L}(d|\overline{\theta})$ is the likelihood of the realisation $d$ with the parameters $\overline{\theta}$, $\pi(\overline{\theta})$ corresponds to our prior on the binary parameters, and $p(d) = \int d \overline{\theta} \mathcal{L}(d|\overline{\theta}) \pi(\overline{\theta})$ is the evidence.
In this work, we assume the so-called zero-noise approximation, i.e. $d = h(\overline{\theta})$ where $h$ denotes the waveform. We refer the interested reader to \citet{Marsat21} for more detail.

The GW signal parameter estimation is performed using the response and likelihood code of \cite{Marsat21}, together with the parallel-tempered ensemble sampler \texttt{ptemcee}~\cite{2016MNRAS.455.1919V} for the Bayesian parameter estimation. We initialize our chains around the simulated signal with a covariance computed from the Fisher matrix, and use enriched proposals that allow to jump to the known potentially degenerate modes in the sky position, inclination and polarization \cite{Marsat21}. 
 For the prior distributions, we assume the sky position angles, inclination and polarization uniformly distributed over the sphere. For all the other parameters we adopt uniform flat priors. 

We run the MCMC chain for each system for $2000$ iterations with 
$64$ walkers \footnote{In the MCMC formalism, a `walker' represents the point of the chain that is exploring the parameter space.} 
and $10$ temperatures \footnote{In the context of parallel tempering, different temperatures are adopted to smooth the parameter space and ease the convergence of the algorithm.}. 
This provides a first set of parameter posteriors. Among them, we are particularly interested in the sky area, to ensure the EM detection. We therefore rerun the chains, for $10^5$ iterations, for all the systems whose the first run produced a posterior with an initial sky localization error of $5< \Delta \Omega/\degsq<40$ at $90\%$ confidence level. 
This way, we ensure the convergence of the MCMC chains for the interesting systems.
Moreover, for all the systems with error in the sky localization $\Delta \Omega < 10 \degsq $ at $90\%$ confidence level, i.e.~those which we use in the rest of the paper, we further check the convergence of the parameter estimation for the parameters $\beta$, $\lambda$ and $d_L$, studying the evolution of the chains as a function of the iterations.
At the end of the process, we obtain three catalogues of EMcps for each of the MBHB astrophysical formation models, for which we are confident that the parameter estimation has converged.


\section{\label{sec:results}Results}

For each of the three MBHB formation astrophysical models described in Sections~\ref{sec:general_strategy} and \ref{sec:MBHB_catalogue}, we use catalogues simulating 90 years of data. 
They contain a total number of 15546, 692 and 10700 MBHBs for Pop3, Q3d and Q3nd respectively. 
The following results are presented for 4 years of LISA observations, i.e.~5 years of mission duration with 80\% duty cycle. 
All catalogues start at $z=20$.

As described in Section \ref{sec:general_strategy}, we first select, among all the events in the catalogues, those with a detectable EM emission, on which we then run the GW signal parameter estimation, to extract the EMcps.
As we shall demonstrate, the number of EMcps strongly depends on how the EM emission is modeled, on the specific instrument adopted to detect it, and on the detection strategy, other than on the astrophysical population model. 
Several choices for how to combine these variables are possible, leading to
different  configurations for the EMcps observations. To simplify the presentation of the results, we focus on two specific models, one maximising the number of EMcps, and one minimising it. 
They are defined as follow:

\begin{itemize}
    \item Maximising
    \begin{itemize}
        \item[-] AGN obscuration neglected
        \item[-] Collimated jet emission with $\Gamma=2$ and isotropic flare
        \item[-] Eddington accretion for X-ray emission
    \end{itemize}
    \item  Minimising 
    \begin{itemize}
        \item[-] AGN obscuration included
        \item[-] Collimated jet and flare emissions with $\Gamma=2$
        \item[-] Catalogue accretion for X-ray emission
    \end{itemize}
\end{itemize}


For these two models, we present the rates of both multimessenger candidates and EMcps. We also discuss possible variations to the two selected models separately. 
Among all the variables, the jet opening angle is the factor that most affects the EMcps rates, while, e.g.,~taking the accretion from the catalogues or assuming it at Eddington does not change the rates significantly. This is because the SKA+ELT combination dominates the rates in the case of an isotropic flare emission, as we will show below.

\subsection{\label{subsec:gen_distro}General distributions}

\begin{table}
\caption{\label{tab:total_event_detected} Average number of intrinsic and GW-detected (i.e., ${\rm SNR}>10$ in LISA) MBHBs in 4 yrs, for different astrophysical models.}
\begin{ruledtabular}
\begin{tabular}{c | m{2.4cm} | m{2.4cm}  } 
   & Total catalogue  & $\SNR>10$  \\
\Xhline{0.1pt}
Pop3   & 690.9  & 129.3  \\
Q3d  & 30.7 & 30.4  \\
Q3nd  & 475.5 & 471.1  \\

\end{tabular}
\end{ruledtabular}

\end{table}

\begin{figure*}
    \includegraphics[width=\textwidth]{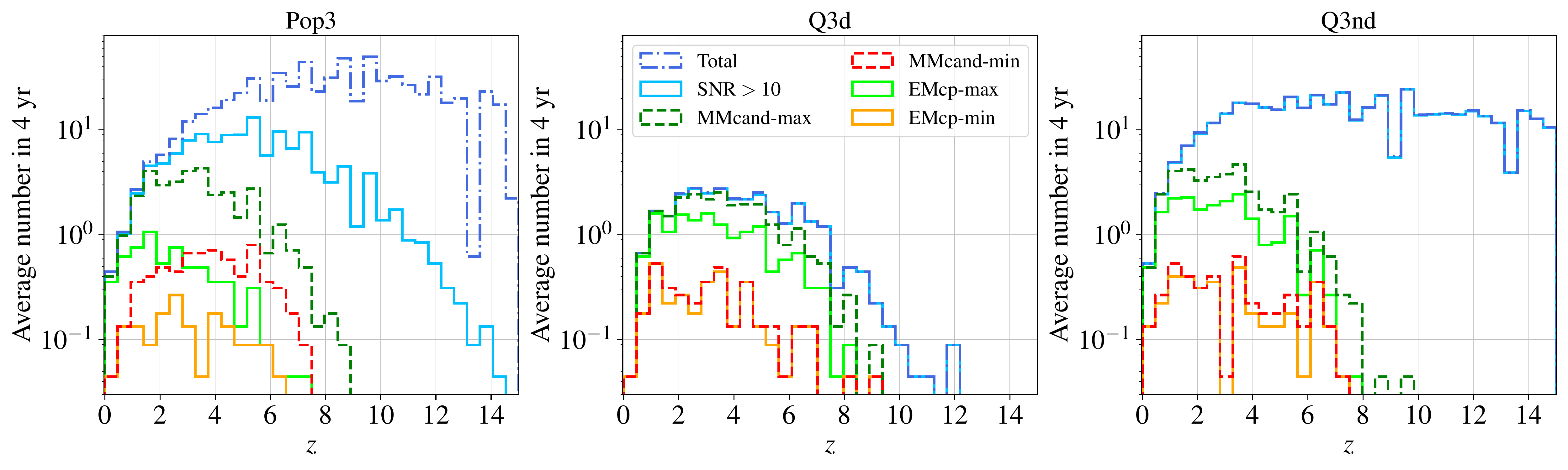} 
    \caption{ Average number of \emph{i)} MBHB mergers directly from the catalogues (dark blue dotted-dashed line), \emph{ii)} MBHB mergers detected in LISA with ${\rm SNR}>10$ (light blue solid line), \emph{iii)} multimessenger candidates (dark green and red dashed lines), and \emph{iv)} EMcps (light green and yellow solid lines), as function of redshift, for the three astrophysical MBHBs formation models, and assuming 4 yrs of LISA observations. 
    In particular, the dark green and the red dashed lines correspond respectively to the multimessenger candidates (c.f.~definition in Section \ref{sec:general_strategy}) in the maximising model, without absorption and isotropic radio flare emission, and in the minimising model, with absorption and $\Gamma =2$ (c.f.~definitions at the beginning of Section \ref{sec:results}). The light green and yellow lines correspond to the EMcps (c.f.~definition in Section \ref{sec:general_strategy}) distributions also in the maximising and minimising models respectively. Applying the requirement of EM detectability and imposing the sky localization threshold select only the closest events, while including absorption and collimated radio emission decreases the overall number of both multimessenger candidates and EMcps.
    }
    \label{fig:z_distro_stsi}
\end{figure*}

\begin{figure*}
    \includegraphics[width=\textwidth]{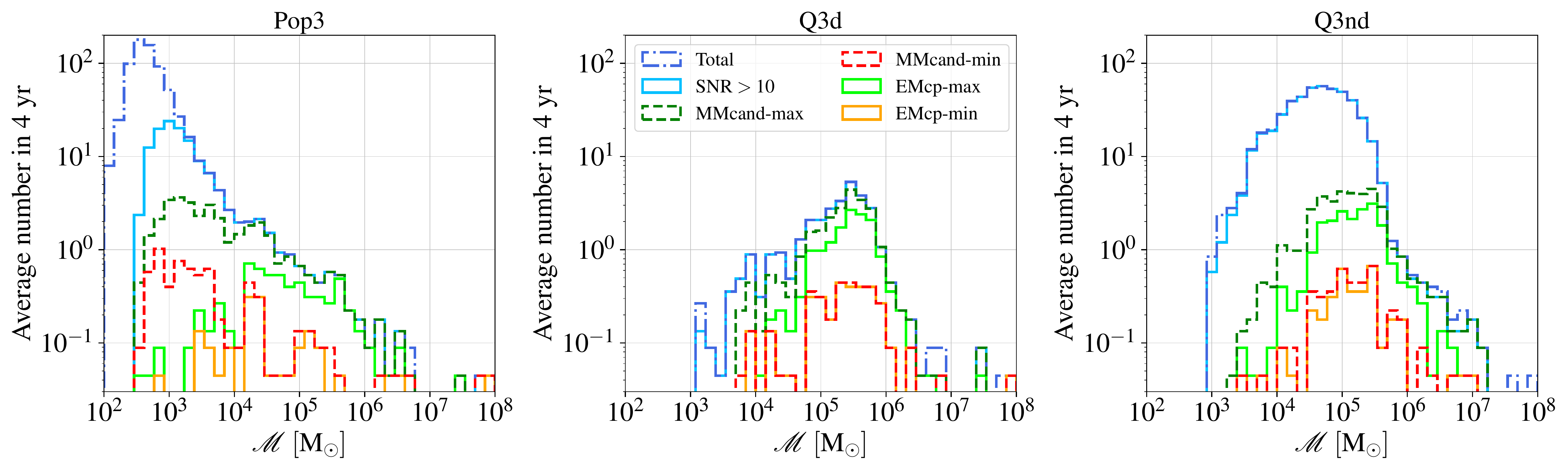} 
    \caption{Same as  Fig.~\ref{fig:z_distro_stsi} as function of chirp mass $\mchirp$. LISA sensitivity selects only systems with $10^4 \lesssim \mchirp/\msun \lesssim 10^6$ as EMcps.}
    \label{fig:mchirp_distro_stsi}
\end{figure*}

\begin{figure*}
    \subfigure{\includegraphics[width=1\textwidth]{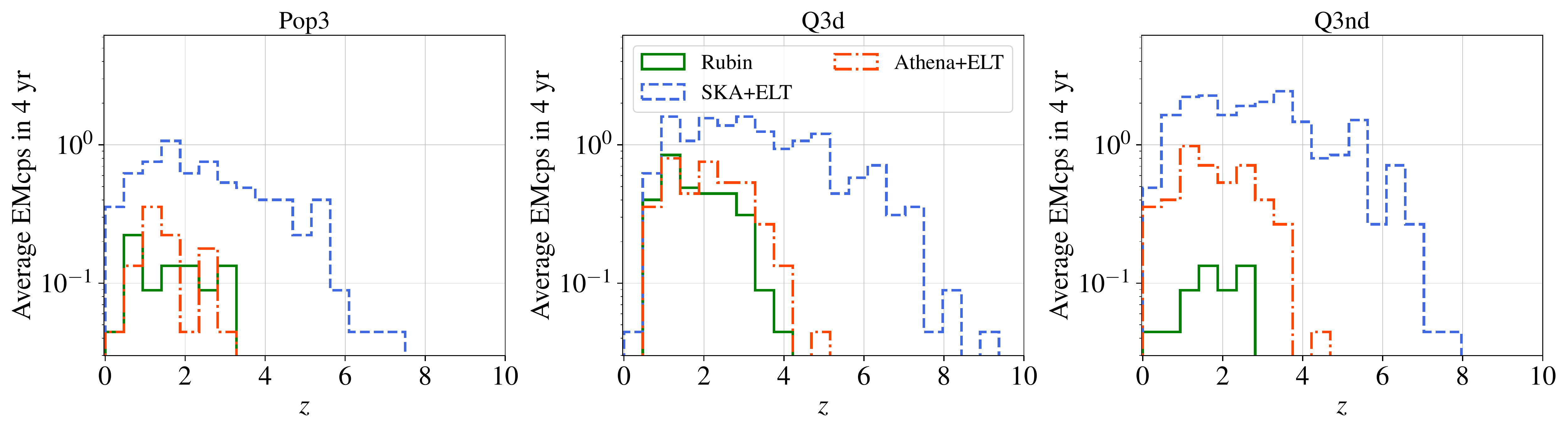}} \\ 
    \vspace{-0.6cm}
    \subfigure{\includegraphics[width=1\textwidth]{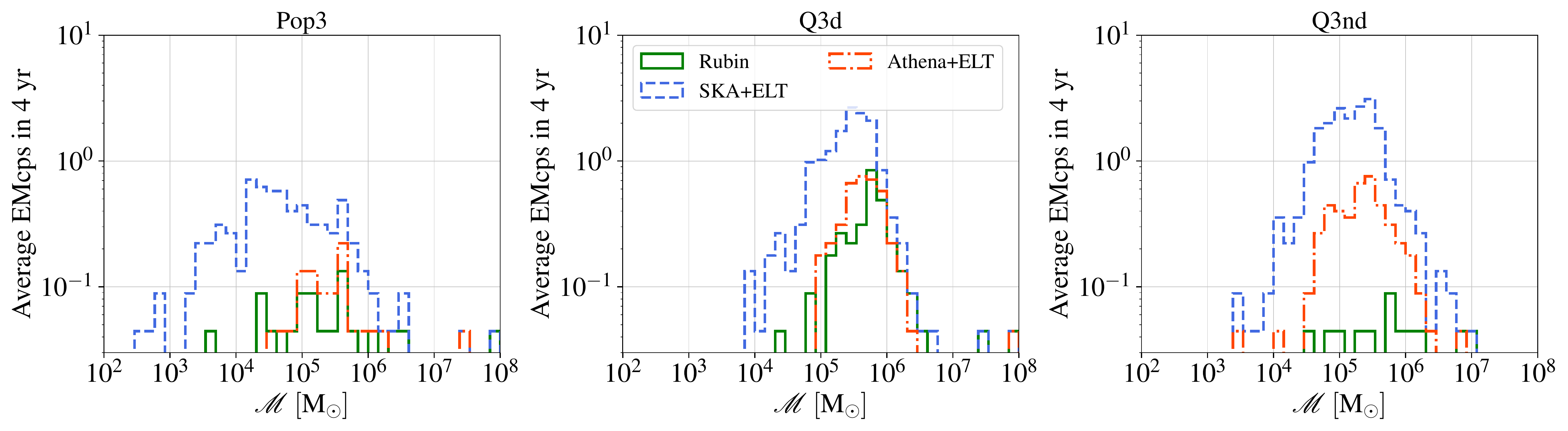}} 
    \caption{Average number of EMcps in each observational scenario, as clarified by the legend, in the maximising case as function of redshift (upper panels) and chirp mass (lower panels) for the three astrophysical models assuming 4 yr of LISA time mission. SKA+ELT provides more EMcps in the maximising scenario thanks to the isotropic flare emission. We stress that we can not simply sum the y-axis for each instrument combination because the same system might be observed by different telescopes at the same time.}
    \label{fig:stsi_distro_3_strategies}
\end{figure*}

In this section we discuss the distributions in redshift and chirp mass of the average number of MBHBs events, multimessenger candidates, and EMcps, in the two models labelled respectively ``maximising" and ``minimising" (c.f.~the beginning of Section \ref{sec:results}).
Furthermore, we also report the average number of multimessenger candidates and EMcps for the other observational scenarios, in Tables \ref{tab:stsi_all_scenarios} and \ref{tab:cp_all_scenarios}.
The average numbers are intended in 4 yrs of observations with LISA, and are obtained by multiplying the total numbers provided by the 90 yrs of catalogues by $4/90$.

In Fig.~\ref{fig:z_distro_stsi} we present the average number of merging binaries as a function of redshift.
Models Pop3 and Q3nd predict a large fraction of mergers at $z\gtrsim10$, while in the Q3d model all the systems merge at $z \lesssim 12$. 
Removing the systems with ${\rm SNR}<10$ in LISA leads to the loss of $\sim80\%$ of high-redshift sources in the Pop3 catalogue, caused by their low mass (see Fig.~\ref{fig:mchirp_distro_stsi}). 
The systems of the Q3nd catalogue are on average more massive, therefore, the SNR cut does not alter their number.
The systems of the Q3d catalogue are also all detected by LISA with ${\rm SNR}>10$: this is expected, since they have a mass distribution similar to Q3nd, and they merge at smaller redshifts. 
The average number of intrinsic and GW-detected events for each of the three astrophysical models is reported in Tab.~\ref{tab:total_event_detected}.

Among the systems with ${\rm SNR}>10$, we further select the multimessenger candidates, i.e.~those with a detectable EM counterpart. 
In Fig.~\ref{fig:z_distro_stsi}, 
we show their distributions in the maximising and minimising models.

\begin{table*}
\caption{\label{tab:cp_all_scenarios} Average number of multimessenger candidates in 4 yrs, in each observational scenario. Fluxes are in units of $\rm erg \, s^{-1} \, cm^{-2} $ (more detail in the text).   }
\begin{tabular}{c|c|c|c|c|c|c|c|c|c } 
\hline \hline 
 \Tstrut & Rubin & \multicolumn{3}{c|}{SKA+ELT} &  \multicolumn{4}{c|}{Athena+ELT} & \\  \cline{3-9}
\multirow{2}{*}{} & \multirow{2}{*}{}  & \multirow{2}{*}{Isotropic flare} & \multirow{2}{*}{$\Gamma 2$} & \multirow{2}{*}{$\Gamma 10$} & \multicolumn{2}{c|}{Catalogue} & \multicolumn{2}{c|}{Eddington} & \\
& & & & & $F_{\rm X, \, lim}$ = 4e-17 & $F_{\rm X, \, lim}$ = 2e-16  &  $F_{\rm X, \, lim}$ = 4e-17 &  $F_{\rm X, \, lim}$ = 2e-16  & \\ 
\Xhline{0.1pt}
\multirow{3}{*}{No-obsc.}  & 1.3 & 34.4 & 6.27 & 0.13 & 2.35 & 0.62 & 3.95 & 1.82 & Pop3\\
&  3.33 & 24.0 & 2.89 & 0.04 & 5.42 & 1.64 & 8.53 & 3.02 & Q3d\\
&  0.84 & 34.5 & 3.78 & 0.04 & 1.6 & 0.44 & 15.9 & 6.31 & Q3nd\\
\Xhline{0.1pt}
\multirow{3}{*}{Obsc.}  & 0.35 & 34.4 & 6.27 & 0.13 & 0.18 & 0.13 & 0.4 & 0.31 & Pop3\\
&  0.8 & 24.0 & 2.89 & 0.04 & 0.35 & 0.17 & 0.53 & 0.13 & Q3d\\
\Bstrut &  0.49 & 34.5 & 3.78 & 0.04 & 0.27 & 0.09 & 1.42 & 0.53 & Q3nd\\

\hline \hline 
\end{tabular}
\end{table*}

\begin{table*}
\caption{\label{tab:stsi_all_scenarios} Average number of EMcps in 4 yrs in each observational scenario. Fluxes are in units of $\rm erg \, s^{-1} \, cm^{-2} $ (more detail in the text). The average number of EMcps obtained combining observations with multiple facilities are reported in Tab.~\ref{tab:stsi_combined_strategies}.   }
\begin{tabular}{c|c|c|c|c|c|c|c|c|c } 
\hline \hline 
 \Tstrut & Rubin & \multicolumn{3}{c|}{SKA+ELT} &  \multicolumn{4}{c|}{Athena+ELT} & \\  \cline{3-9}
\multirow{2}{*}{} & \multirow{2}{*}{}  & \multirow{2}{*}{Isotropic flare} & \multirow{2}{*}{$\Gamma 2$} & \multirow{2}{*}{$\Gamma 10$} & \multicolumn{2}{c|}{Catalogue} & \multicolumn{2}{c|}{Eddington} &  \\
& & & & & $F_{\rm X, \, lim}$ = 4e-17 & $F_{\rm X, \, lim}$ = 2e-16  &  $F_{\rm X, \, lim} $ = 4e-17 &  $F_{\rm X, \, lim} $ = 2e-16 &  \\ \cline{2-5}
& \multicolumn{4}{c|}{} & & & & &\\[-0.9em]
& \multicolumn{4}{c|}{$\Delta \Omega = 10 \degsq$} & $\Delta \Omega = 0.4 \degsq$ & $\Delta \Omega = 2 \degsq$ & $\Delta \Omega = 0.4 \degsq$ & $\Delta \Omega = 2 \degsq$ & \\
\Xhline{0.1pt}
\multirow{3}{*}{No-obsc.}  & 0.84 & 6.4 & 1.51 & 0.04 & 0.49 & 0.27 & 1.02 & 0.84 & Pop3\\
&  3.07 & 14.8 & 2.71 & 0.04 & 2.67 & 1.38 & 3.87 & 2.13 & Q3d \\
&  0.53 & 20.3 & 3.2 & 0.04 & 0.58 & 0.31 & 4.4 & 3.24 & Q3nd\\
\Xhline{0.1pt}
\multirow{3}{*}{Obsc.}  & 0.13 & 6.4 & 1.51 & 0.04 & 0.04 & 0.04 & 0.13 & 0.17 & Pop3 \\
&  0.75 & 14.8 & 2.71 & 0.04 & 0.22 & 0.13 & 0.18 & 0.09 & Q3d\\
\Bstrut &  0.35 & 20.3 & 3.2 & 0.04 & 0.18 & 0.04 & 0.27 & 0.31 & Q3nd \\

\hline \hline 
\end{tabular}
\end{table*}

The additional requirement of  EM detectability selects  systems at even smaller redshift: for all the three astrophysical scenarios, multimessenger candidates have $z<10$. 
Within the maximising model, we predict in total $24.0$ ($35$) \{$37.6$\} multimessenger candidates for Q3d (Pop3) \{Q3nd\} in 4 years.
As expected, if we include obscuration and collimated radio emission, the multimessenger candidates number decreases to $3.6$ ($6.6$) \{$4.2$\}, and only  systems at $z\lesssim 8$ can be detected. 
This reduction of about $ 81-89\%$ in the number of multimessenger candidates with respect to the maximising model, is similar in all astrophysical models. 

At last, we impose a cut in the sky localization of the systems, to select only the EMcps. 
We obtain $14.8$ ($6.4$) \{$20.7$\} EMcps  for Q3d (Pop3) \{Q3nd\} in 4 yrs in the maximising model, and nothing but $3.3$ ($1.6$) \{$3.5$\}  if we include AGN \obsc and collimated flare and jet emission (minimising model).
The Pop3 scenario predicts the largest number of multimessenger candidates, however, only $\sim20\%$ among them are promoted to  EMcps: LISA will not localize these sources accurately enough, due to their intrinsic low chirp mass and high redshift. 
On the other hand, the Q3d and Q3nd models predict fewer multimessenger candidates, but $\sim 60\%$ among them are EMcps. 
Within the minimising model, even though the total number of both multimessenger candidates and EMcps decreases,
the fraction of multimessenger candidates promoted to EMcps is higher for all astrophysical models: $24\%$, $92\%$, and $83\%$ for Pop3, Q3d and Q3nd respectively, as opposed to $18\%$, $62\%$, and $55\%$ in the maximising model. 
We interpret this fact as follows: including obscuration and collimated radio emission effectively removes the tails of the distributions, selecting the bulk of the ``best'' events: those with redshift low enough to have good LISA parameter estimation, but high enough to be sufficiently numerous. Further imposing the sky localization cut has therefore a minor effect on this subset of events.

\begin{table*}
\caption{\label{tab:stsi_combined_strategies} Average number of EMcps in 4 yrs combining the different observational models. The case \emph{without obscuration} (\emph{with obscuration}) is reported in the left (right) table. Fluxes are in units of $\rm erg \, s^{-1} \, cm^{-2} $. The EMcps rates corresponding to the \emph{maximising} (\emph{minimising}) case are in boldface in the left (right) table.  }
\resizebox{\columnwidth}{!}{
\begin{tabular}{c|c|c|c|c|c } 
\hline \hline 
 \Tstrut  & SKA & \multicolumn{2}{c|}{Athena} & &\\
  & $\Delta \Omega = 10 \degsq$ & \multicolumn{2}{c|}{No obsc.} & &\\
 \cline{2-5}
 \multirow{36}{*}{\begin{tabular}{@{}c@{}}Rubin \\ $\Delta \Omega = 10 \degsq$ \\ No obsc. \end{tabular}} & \multirow{12}{*}{\begin{tabular}{@{}c@{}}Isotropic \\ flare \end{tabular}} & \multirow{6}{*}{Catalogue} & \multirow{3}{*}{\begin{tabular}{@{}c@{}}$F_{\rm X, \, lim}$ = 4e-17 \\ $\Delta \Omega = 0.4 \degsq$ \end{tabular}} & 6.4 & \multirow{36}{*}{\begin{tabular}{@{}c@{}}Pop3 \\ Q3d \\ Q3nd \end{tabular}} \\
 & & & & 14.8 &  \\
 & & & & 20.4 & \\  \cline{4-5}
 & & & \multirow{3}{*}{\begin{tabular}{@{}c@{}}$F_{\rm X, \, lim}$ = 2e-16 \\ $\Delta \Omega = 2 \degsq$ \end{tabular}} & 6.4 & \\
 & & & & 14.8 & \\
 & & & & 20.4 & \\  \cline{3-5}
 & & \multirow{6}{*}{Eddington} & \multirow{3}{*}{\begin{tabular}{@{}c@{}}$F_{\rm X, \, lim}$ = 4e-17 \\ $\Delta \Omega = 0.4 \degsq$ \end{tabular}} & {\bf6.4}  & \\
 & & & & {\bf14.8} & \\
 & & & & {\bf20.7} & \\ \cline{4-5}
 & & & \multirow{3}{*}{\begin{tabular}{@{}c@{}}$F_{\rm X, \, lim}$ = 2e-16 \\ $\Delta \Omega = 2 \degsq$ \end{tabular}} & 6.4 & \\
 & & & & 14.8 &\\
 & & & & 20.6 &\\ \cline{2-5}
 & \multirow{12}{*}{$\Gamma2$} & \multirow{6}{*}{Catalogue} & \multirow{3}{*}{\begin{tabular}{@{}c@{}}$F_{\rm X, \, lim}$ = 4e-17 \\ $\Delta \Omega = 0.4 \degsq$ \end{tabular}} & 2.31 &\\
 & & & & 6.18 &\\
 & & & & 3.9 &\\ \cline{4-5}
 & & & \multirow{3}{*}{\begin{tabular}{@{}c@{}}$F_{\rm X, \, lim}$ = 2e-16 \\ $\Delta \Omega = 2 \degsq$ \end{tabular}} & 2.18 &\\
 & & & & 5.5 &\\
 & & & & 3.6 &\\ \cline{3-5}
 & & \multirow{6}{*}{Eddington} & \multirow{3}{*}{\begin{tabular}{@{}c@{}}$F_{\rm X, \, lim}$ = 4e-17 \\ $\Delta \Omega = 0.4 \degsq$ \end{tabular}} & 2.8 & \\
 & & & & 7.0 & \\
 & & & & 6.9 & \\ \cline{4-5}
 & & & \multirow{3}{*}{\begin{tabular}{@{}c@{}}$F_{\rm X, \, lim}$ = 2e-16 \\ $\Delta \Omega = 2 \degsq$ \end{tabular}} & 2.67 &\\
 & & & & 6.0 &\\
 & & & & 5.9 & \\ \cline{2-5}
 & \multirow{12}{*}{$\Gamma10$} & \multirow{6}{*}{Catalogue} & \multirow{3}{*}{\begin{tabular}{@{}c@{}}$F_{\rm X, \, lim}$ = 4e-17 \\ $\Delta \Omega = 0.4 \degsq$ \end{tabular}} & 1.07 &\\
 & & & & 4.04 & \\
 & & & & 0.9 &\\ \cline{4-5}
 & & & \multirow{3}{*}{\begin{tabular}{@{}c@{}}$F_{\rm X, \, lim}$ = 2e-16 \\ $\Delta \Omega = 2 \degsq$ \end{tabular}} & 0.9 &\\
 & & & & 3.2 &\\
 & & & & 0.58 & \\ \cline{3-5}
 & & \multirow{6}{*}{Eddington} & \multirow{3}{*}{\begin{tabular}{@{}c@{}}$F_{\rm X, \, lim}$ = 4e-17 \\ $\Delta \Omega = 0.4 \degsq$ \end{tabular}} & 1.6 &\\
 & & & & 5.2 &\\
 & & & & 4.7 &\\ \cline{4-5}
 & & & \multirow{3}{*}{\begin{tabular}{@{}c@{}}$F_{\rm X, \, lim}$ = 2e-16 \\ $\Delta \Omega = 2 \degsq$ \end{tabular}} & 1.5 &\\
 & & & & 3.9  & \\
 & & & & 3.5  & \\ \cline{2-5}
\hline \hline 
\end{tabular}
}
\resizebox{0.985\columnwidth}{!}{
\begin{tabular}{c|c|c|c|c|c } 
\hline \hline 
  \Tstrut  & SKA & \multicolumn{2}{c|}{Athena} & &\\
  & $\Delta \Omega = 10 \degsq$ & \multicolumn{2}{c|}{with obsc.} & &\\
 \cline{2-5}
\multirow{36}{*}{\begin{tabular}{@{}c@{}}Rubin \\ $\Delta \Omega = 10 \degsq$ \\ with obsc. \end{tabular}} & \multirow{12}{*}{\begin{tabular}{@{}c@{}}Isotropic \\ flare \end{tabular}} & \multirow{6}{*}{Catalogue} & \multirow{3}{*}{\begin{tabular}{@{}c@{}}$F_{\rm X, \, lim}$ = 4e-17 \\ $\Delta \Omega = 0.4 \degsq$ \end{tabular}} & 6.4 & \multirow{36}{*}{\begin{tabular}{@{}c@{}}Pop3 \\ Q3d \\ Q3nd \end{tabular}} \\
 & & & & 14.8 & \\
 & & & & 20.4 & \\  \cline{4-5}
 & & & \multirow{3}{*}{\begin{tabular}{@{}c@{}}$F_{\rm X, \, lim}$ = 2e-16 \\ $\Delta \Omega = 2 \degsq$ \end{tabular}} & 6.4 &\\
 & & & & 14.8 &  \\
 & & & & 20.4 & \\  \cline{3-5}
 & & \multirow{6}{*}{Eddington} & \multirow{3}{*}{\begin{tabular}{@{}c@{}}$F_{\rm X, \, lim}$ = 4e-17 \\ $\Delta \Omega = 0.4 \degsq$ \end{tabular}} & 6.4 & \\
 & & & & 14.8 & \\
 & & & & 20.4 & \\ \cline{4-5}
 & & & \multirow{3}{*}{\begin{tabular}{@{}c@{}}$F_{\rm X, \, lim}$ = 2e-16 \\ $\Delta \Omega = 2 \degsq$ \end{tabular}} & 6.4 &\\
 & & & & 14.8 &  \\
 & & & & 20.4 & \\ \cline{2-5}
 & \multirow{12}{*}{$\Gamma2$} & \multirow{6}{*}{Catalogue} & \multirow{3}{*}{\begin{tabular}{@{}c@{}}$F_{\rm X, \, lim}$ = 4e-17 \\ $\Delta \Omega = 0.4 \degsq$ \end{tabular}} & {\bf1.6} &\\
 & & & & {\bf3.3} &  \\
 & & & & {\bf3.5} & \\ \cline{4-5}
 & & & \multirow{3}{*}{\begin{tabular}{@{}c@{}}$F_{\rm X, \, lim}$ = 2e-16 \\ $\Delta \Omega = 2 \degsq$ \end{tabular}} & 1.6 & \\
 & & & & 3.3 &  \\
 & & & & 3.5  & \\ \cline{3-5}
 & & \multirow{6}{*}{Eddington} & \multirow{3}{*}{\begin{tabular}{@{}c@{}}$F_{\rm X, \, lim}$ = 4e-17 \\ $\Delta \Omega = 0.4 \degsq$ \end{tabular}} & 1.8 & \\
 & & & & 3.5 & \\
 & & & & 3.6 & \\ \cline{4-5}
 & & & \multirow{3}{*}{\begin{tabular}{@{}c@{}}$F_{\rm X, \, lim}$ = 2e-16 \\ $\Delta \Omega = 2 \degsq$ \end{tabular}} & 1.8 &\\
 & & & & 3.4 &  \\
 & & & & 3.7  & \\ \cline{2-5}
 & \multirow{12}{*}{$\Gamma10$} & \multirow{6}{*}{Catalogue} & \multirow{3}{*}{\begin{tabular}{@{}c@{}}$F_{\rm X, \, lim}$ = 4e-17 \\ $\Delta \Omega = 0.4 \degsq$ \end{tabular}} & 0.18 &\\
 & & & & 0.80 & \\
 & & & & 0.49 & \\ \cline{4-5}
 & & & \multirow{3}{*}{\begin{tabular}{@{}c@{}}$F_{\rm X, \, lim}$ = 2e-16 \\ $\Delta \Omega = 2 \degsq$ \end{tabular}} & 0.18 &\\
 & & & & 0.80 & \\
 & & & & 0.40 & \\ \cline{3-5}
 & & \multirow{6}{*}{Eddington} & \multirow{3}{*}{\begin{tabular}{@{}c@{}}$F_{\rm X, \, lim}$ = 4e-17 \\ $\Delta \Omega = 0.4 \degsq$ \end{tabular}} & 0.31 &\\
 & & & & 0.98 &\\
 & & & & 0.67 & \\ \cline{4-5}
 & & & \multirow{3}{*}{\begin{tabular}{@{}c@{}}$F_{\rm X, \, lim}$ = 2e-16 \\ $\Delta \Omega = 2 \degsq$ \end{tabular}} & 0.35 &\\
 & & & & 0.84 \\
 & & & & 0.71 \\ \cline{2-5}
\hline \hline 
\end{tabular}
}
\end{table*}

In Fig.~\ref{fig:mchirp_distro_stsi} we report the same quantities as a function of the rest-frame chirp mass $\mchirp$. 
While the distributions in the massive models (Q3d and Q3nd)  peak  at $\mchirp \sim 10^5-10^6 \msun$, for the Pop3 model the peak of the distribution is at $\mchirp \lesssim 10^3 \msun$, due to the different BH formation processes. 
The $\SNR>10$ cut therefore operates similarly to what already observed for the redshift distribution, i.e. it excludes low-mass events in the Pop3 scenario while leaving the Q3d and Q3nd practically unaffected.

In the maximising model, all the systems with $\mchirp \gtrsim 10^5 \msun$ have  detectable EM emission, while at lower masses a significant fraction of Pop3 and Q3nd can be detected by LISA with $\SNR>10$ but do not have observable EM emission due to the low BH mass or high redshift of the systems.
Adding the further requirement on the sky localization results in an overall rescaling of the multimessenger candidate distributions for the massive astrophysical models, while it selects only the heaviest binaries in the Pop3 model.
As already observed for the distributions as a function of redshift, the reduction in the number of events when one includes obscuration and the collimated jet is higher than the one obtained when one imposes the cut in the sky localization.

In Fig.~\ref{fig:stsi_distro_3_strategies} we show the EMcps distributions for the Rubin Observatory, SKA+ELT and Athena+ELT strategies separately in the maximising scenario. SKA+ELT is the only combination to provide EMcps at $z\gtrsim4$ while the Rubin Observatory and Athena+ELT can observe only the closest events with the latter reaching slightly higher redshifts than the former. 
Moreover we note that we have no detections with the Rubin Observatory above $z>4$ so we can safely use the $r$ band without worrying about absorption. Moving to the chirp mass, the SKA+ELT scenario is able to probe the lightest systems in our catalogues, especially for Pop3, while Athena+ELT and the Rubin Observatory detect the EM emission from systems with $10^4 < \mchirp / M_{\odot} < 10^6$.

The average number of multimessenger candidates and EMcps for each observational scenario is reported in Tab.~\ref{tab:cp_all_scenarios} and Tab.~\ref{tab:stsi_all_scenarios}.
Overall, the observational strategy providing the most multimessenger candidates and EMcps in 4 yrs is  SKA+ELT, if we assume that the radio flare emission is isotropic. 
Accounting for a beamed emission with $\Gamma=2$ provides numbers which  are closer to those obtained when observing with the Rubin Observatory, or with Athena+ELT. 
If we further decrease the opening angle ($\Gamma=10$), observations with SKA+ELT become irrelevant. 
While the beamed emission allows us to detect systems that are farther away from us, imposing that the observer has to be on-axis excludes the vast majority of the systems.

Observing with the Rubin Observatory provides $\sim 3$ EMcps in 4 yrs in the Q3d model without accounting for obscuration, while in all the other astrophysical cases the rates  are below 1. 
Concerning the combination Athena+ELT, as expected the Eddington accretion leads to more multimessenger candidates and more EMcps, since the EM emission is brighter. Moreover, in general the observational strategy where one observes a single region of $\Delta \Omega = 0.4 \degsq$ allows for the detection of slightly more EMcps than the strategy in which one observes a region of $\Delta \Omega = 2 \degsq$ at a higher flux threshold, because there are more systems at fainter fluxes compared to systems with poorer localization.

In general, the two models with massive progenitors predict more EMcps than the Pop3 one, due to the aforementioned difficulties in localising light events with LISA. Indeed, one can appreciate that the Pop3 astrophysical formation model leads to more multimessenger candidates than the Q3nd one, when observing with the Rubin Observatory and SKA+ELT. However, most of them do not satisfy the sky localization requirement and consequently are not accounted for as EMcps.

We highlight that, in order to get the total average number of multimessenger candidates and/or EMcps, one should combine the different EM facilities, while taking care not to count the same event twice (since the same system can be detected with different instruments - see following paragraph). 
The total average number of EMcps is reported in Tab.~\ref{tab:stsi_combined_strategies}.

In Tab.~\ref{tab:number_emcps_observed_with_same_instruments} we report the number of EMcps that can be observed simultaneously by: (i) the Rubin Observatory and SKA (`Rubin+SKA');  (ii) the Rubin Observatory and Athena (`Rubin+Athena'); (iii) SKA and Athena (`SKA+Athena'); (iv) the three instruments (`All'). In the maximising scenario, the Q3d model predicts $\sim 2-4$ EMcps in 4 yr (depending on the instruments considered), and about $\sim2$ EMcps should be observable by all the instruments simultaneously. As expected, the combination SKA+Athena provides the largest numbers, since both SKA and Athena can observe sources at higher redshift than the Rubin Observatory (c.f.~Fig.~\ref{fig:stsi_distro_3_strategies}). Moving to the minimising case, we find that  $<0.3$ EMcps in 4 yrs can be observed by multiple instruments simultaneously, regardless of the astrophysical model.     
\begin{table}
\caption{\label{tab:number_emcps_observed_with_same_instruments}Average number of EMcps observed simultaneously by multiple instruments, in 4 yrs and for different combinations. Upper (lower) table refers to the maximising (minimising) scenario. }
\begin{ruledtabular}
\begin{tabular}{c|c|c|c|c} 
& \multicolumn{4}{c}{Maximising (multiple instruments)} \\
\hline  
 & Rubin+SKA & Rubin+Athena & SKA+Athena & All \\
\hline  
Pop3   & 0.84  & 0.31  & 1.02  & 0.31    \\
Q3d  & 3.07 & 1.73 & 3.9  & 1.7    \\
Q3nd  & 0.5 & 0.27 & 4.0  & 0.22   \\
\end{tabular}

\bigskip 

\begin{tabular}{c|c|c|c|c} 
& \multicolumn{4}{c}{Minimising (multiple instruments)} \\
\hline  
 & Rubin+SKA & Rubin+Athena & SKA+Athena & All \\
\hline  
Pop3   & $<$0.04  & 0.04  & $<$0.04  & $<$0.04    \\
Q3d  & 0.13 & 0.22 & 0.13  & 0.13    \\
Q3nd  & 0.09 & 0.09 & 0.04  & $<$0.04   \\
\end{tabular}
\end{ruledtabular}
\end{table}

\subsection{\label{subsec:mmcand_and_emcp_without_redshift}MMcands and EMcps without redshift measurement }

\begin{table}
\caption{\label{tab:number_ska_alone}Average number of MMcands and EMcps in 4 yrs, for different astrophysical models for the scenario SKA alone. For the Athena only case we obtain the exact same numbers reported in Tab.~\ref{tab:cp_all_scenarios} and Tab.~\ref{tab:stsi_all_scenarios}. }
\begin{ruledtabular}
\begin{tabular}{c|c|c|c|c|c|c} 
 & \multicolumn{3}{c|}{MMcands} & \multicolumn{3}{c}{EMcps} \\\hline  

   &  \begin{tabular}{@{}c@{}}Isotropic \\ flare \end{tabular} &  $\Gamma 2$  &  $\Gamma 10$ &  \begin{tabular}{@{}c@{}}Isotropic \\ flare \end{tabular}  &  $\Gamma 2$  &  $\Gamma 10$ \\
   
\hline  
Pop3   & 85.8  & 26.8  & 0.58  & 6.5  & 1.51  & 0.04  \\
Q3d  & 29.3 & 3.4 & 0.04  & 15.4  & 2.84 & 0.04  \\
Q3nd  & 125 & 18.0 & 0.93  & 24.1  & 4.13  & 0.13 \\
\end{tabular}
\end{ruledtabular}
\end{table}

In this section we relax the requirement of the redshift determination, i.e. we present the predicted number of multimessenger candidates and EMcps, but without imposing that their redshift should be measured independently. Indeed, interesting information on how the radio or X-ray emissions are produced can also be inferred exclusively by the detection of the EM emission. The redshift can then be determined from the GW-measured luminosity distance by assuming the standard model cosmology (it will not be possible, though, to use these EMcps as standard sirens).
Relaxing the redshift determination requirement does not change significantly the number of EMcps; it only affects the number of MMcands, which are, however, less interesting because their sky localization is unknown. In the following, the requirements for the identification of the MMcands and EMcps in terms of $\SNR$, flux and sky localization remain the same as in the rest of the paper.

First, the number of MMcands and EMcps detectable with the Rubin Observatory with and without imposing redshift determination remains the same because the threshold magnitude that we adopted for spectroscopy, $m_{\rm AGN, Rubin, \, lim} = 27.5$, is close to the photometric limit of the survey. 

Second, let us focus on the MMcands and EMcps detectable with Athena only. If redshift is not needed, this amounts to dropping the requirement of detectability with ELT, which was imposed exclusively for the redshift determination.  However, as can be appreciated from Fig.~\ref{fig:stsi_distro_3_strategies}, Athena can only detect sources up to $z \lesssim 4$ while ELT is sensitive up to $z \lesssim 8$ (we justify this value confronting the results for the multimessenger candidates cases in Fig.~\ref{fig:z_distro_stsi} with the 'SKA alone' configuration in Fig.~\ref{fig:mmcand_distro_ska_only} - see discussion at the end of this subsection ) . Therefore all the sources detectable with Athena can also be observed with ELT, so that the number of MMcands and EMcps with and without redshift determination is identical.

Moving to SKA, removing the requirement on the redshift determination increases the number of MMcands for Pop3 and Q3nd, respectively, by a factor $\sim 2.5$ and $\sim 3.6$ in the isotropic flare scenario. In the models with beamed emissions, the ratio between the number of MMcands without  and with redshift rises to $\sim 4.3$ and $4.8$ for Pop3 and Q3nd respectively in the case $\Gamma2$ and to $\sim 4.5$ and $\sim 23 $ in the $\Gamma 10$ scenario. For Q3d the increase is less significant in all configurations.  The limiting factor is that ELT reaches lower redshifts than SKA, as can be appreciated comparing  Fig.~\ref{fig:z_distro_stsi} and Fig.~\ref{fig:mmcand_distro_ska_only}. The number of EMcps does not change significantly instead, because the requirement on the sky localization selects lower redshift systems which can be detected by ELT. 

In Tab.~\ref{tab:number_ska_alone} we present the numbers of MMcands and EMcps observable with SKA alone, to be compared with the values reported in Tab.~\ref{tab:cp_all_scenarios} and Tab.~\ref{tab:stsi_all_scenarios} which include detection with ELT. The distributions of MMcands for SKA in terms of redshift and chirp mass are reported in Fig.~\ref{fig:mmcand_distro_ska_only}.

\subsection{\label{subsec:magn_flux_distro}Magnitudes and fluxes distributions for EMcps}

\begin{figure}
    \includegraphics[width=0.5\textwidth]{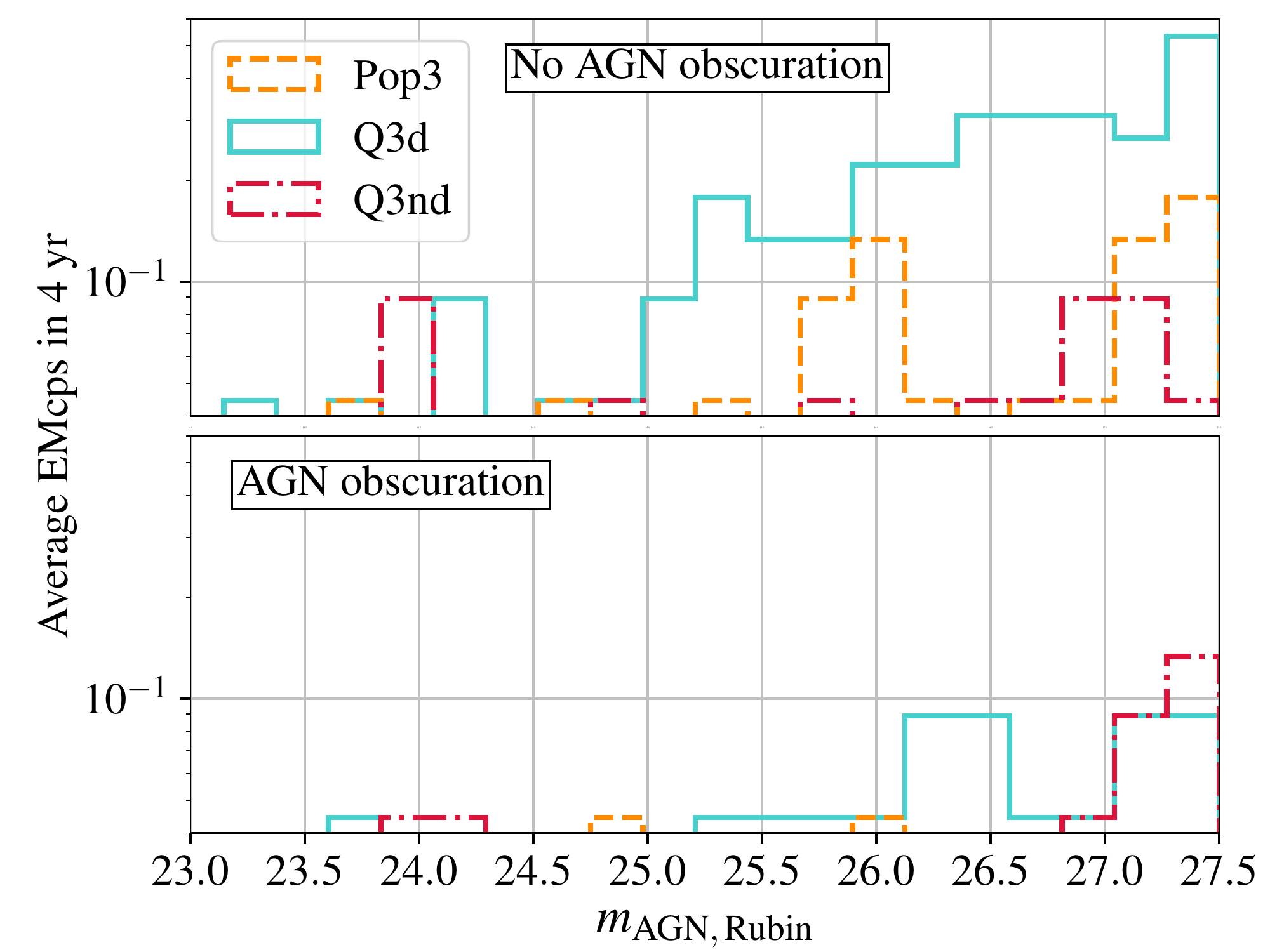}
    \caption{Magnitude distributions for the EMcps  detected with the Rubin Observatory for the three astrophysical scenarios, as clarified by the legend. The upper (lower) panel corresponds to the case without (with) AGN obscuration.}
    \label{fig:magn_lsst_distro_stsi}
\end{figure}

\begin{figure}
    \includegraphics[width=0.5\textwidth]{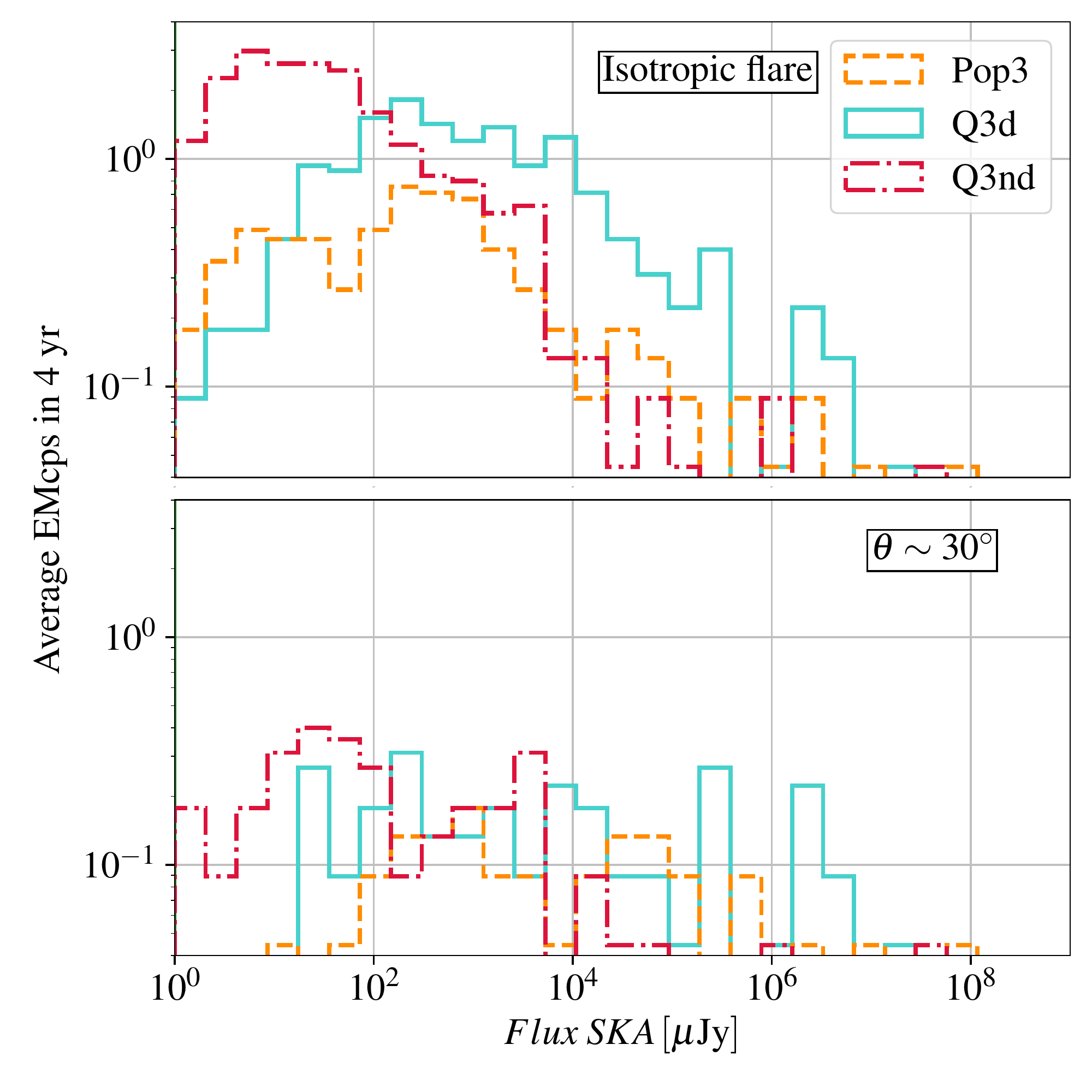} 
    \caption{Flux distributions for the EMcps detected with the SKA+ELT configuration  for the three astrophysical scenarios, as clarified by the legend. The upper (lower) panel corresponds to the case with isotropic flare ($\Gamma =2$ jet) radio emission.}
    \label{fig:magn_ska_distro_stsi}
\end{figure}

In this section we present the average number of EMcps as a function of magnitude (relevant for detection with the Rubin Observatory and ELT) and flux (relevant for detection with SKA and Athena).

The magnitude distributions for the EMcps detectable with the Rubin Observatory are reported in Fig.~\ref{fig:magn_lsst_distro_stsi}. 
In the absence of AGN obscuration, most of the systems have $m_{\rm AGN, Rubin} >25 $  and accumulate toward higher magnitudes. From the distribution it is also evident that the Q3d model provides the highest average number of EMcps compared to Pop3 and Q3nd.
If we account for AGN obscuration, the number of EMcps diminishes appreciably and the typical magnitude increases to $m_{\rm AGN, Rubin} \gtrsim 26 $,  while Q3d still remains the most promising scenario.

The distributions of the radio fluxes of the EMcps detectable with SKA+ELT  are reported in Fig.~\ref{fig:magn_ska_distro_stsi} for the `isotropic flare' and `$\Gamma2$' scenarios.
In the isotropic flare case, the distributions are characterised by a peak around $\lesssim 10^2 \mu \rm Jy$, for all astrophysical models. In the $\Gamma2$ case, the distributions appear instead to be flatter. 

By inspecting how the radio luminosities are distributed in the catalogues (c.f.~Fig.~\ref{fig:Lradio_fluxes_distribution} in Appendix \ref{app:useful_plot}) we found that the flare emission occurs typically at lower luminosity than the jet one: it is therefore subdominant with respect to the jet emission. 
However, the jet is pointing in the direction of the observer only in a small fraction of cases. 
Therefore, the peak in the distribution observed in the `isotropic flare scenario' corresponds to the flare emission. 
In the `$\Gamma2$' scenario, on the other hand, the jet luminosity dominates, and the peak characteristic of the isotropic flare scenario is absent.

\begin{figure*}
    \includegraphics[width=\textwidth]{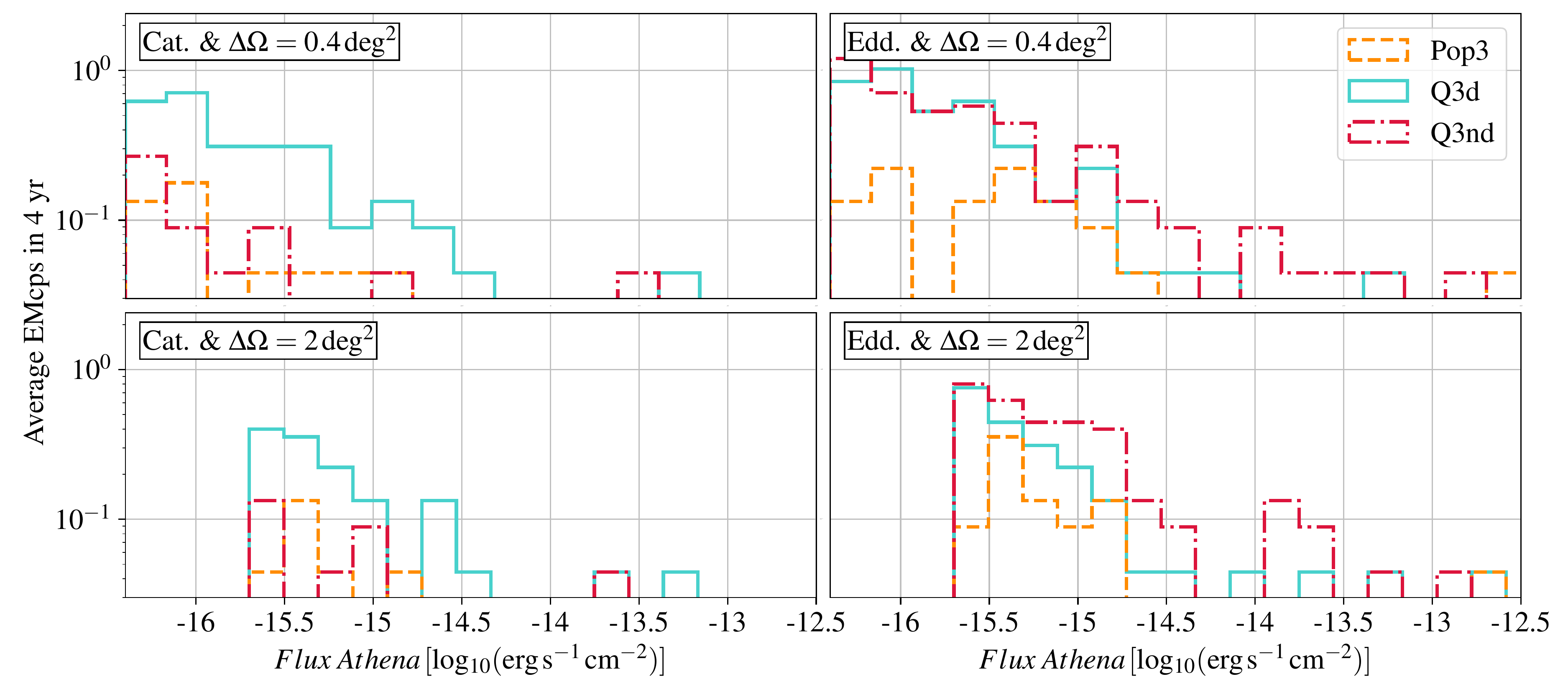} 
    \caption{Flux distributions for the EMcps detected with the Athena+ELT configuration  for the three astrophysical scenarios and different configurations.}
    \label{fig:magn_athena_distro_stsi}
\end{figure*}

\begin{figure}
    \includegraphics[width=0.5\textwidth]{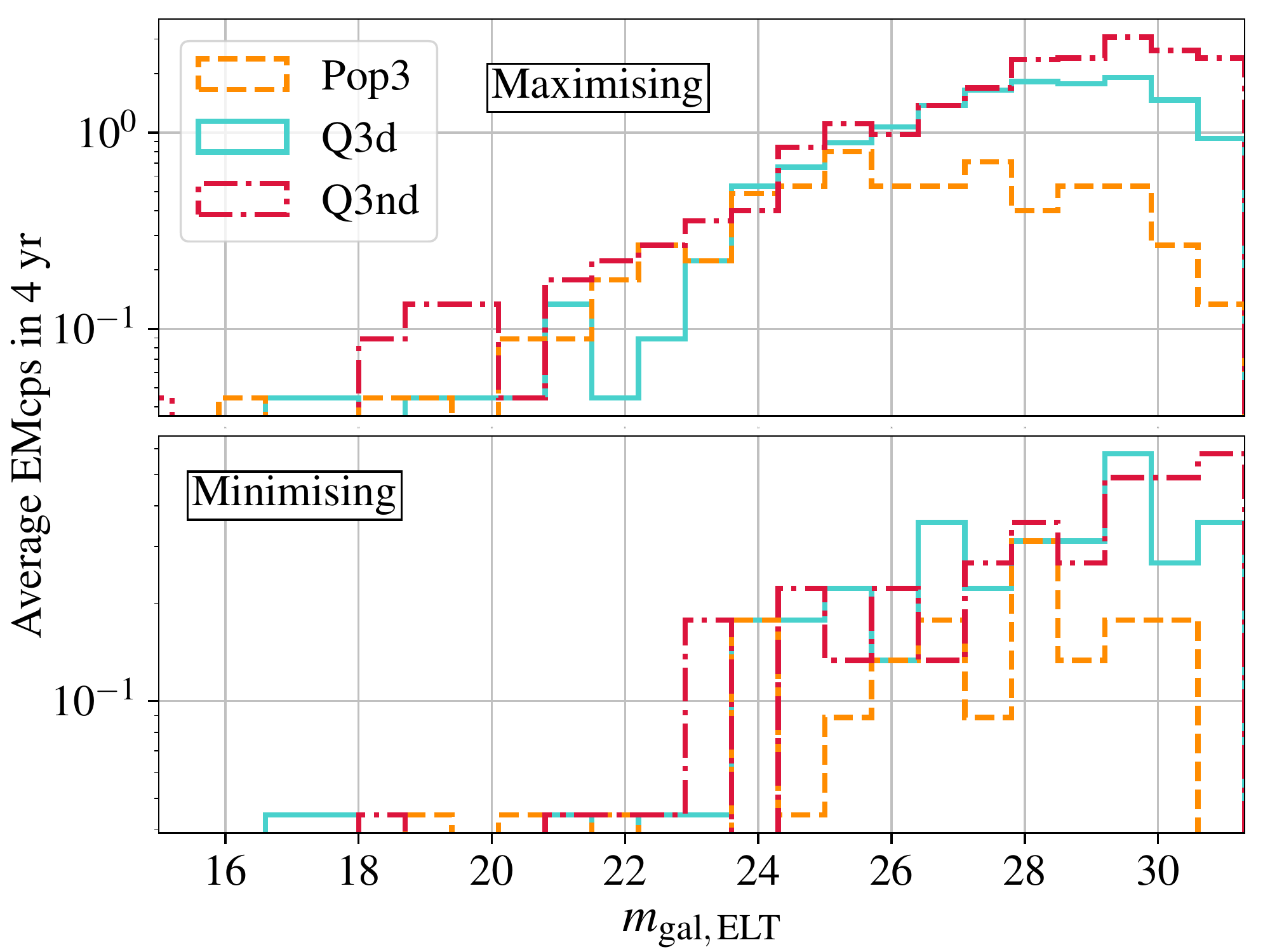} 
    \caption{Magnitude distributions for the EMcps observable with the SKA+ELT and Athena+ELT. The upper (lower) panel corresponds the maximising (minimising) scenario.}
    \label{fig:magn_elt_distro_stsi}
\end{figure}

In Fig.~\ref{fig:magn_athena_distro_stsi} we report the number of EMcps observable with Athena as a function of  their X-ray flux. We account both for accretion evaluated from the amount of gas surrounding the MBHB (estimated by the SAM), or at Eddington. Furthermore, we show the two sky localization thresholds, within the corresponding flux limits.
If we consider the scenario where the accretion rate is derived from the catalogue, the Q3d model provides the highest EMcps number, while Pop3 and Q3nd give similar results. From Fig.~\ref{fig:Ledd_ratio_vs_mchirp} in Appendix \ref{app:useful_plot}, it can be appreciated that, in the Pop3 scenario,  the Eddington ratio is about the same as in the Q3d scenario for systems with $\mchirp >10^4-10^5 \msun$, however, the number of systems with high mass is intrinsically lower with respect to Q3d, and therefore the overall EMcps rate is lower. Furthermore, in Q3nd there are overall more events, but the Eddington ratio  is reduced (in other words, without delays  there is not time to accumulate enough gas to be accreted on the binary) which leads to fewer EMcps. 
 In the entire catalogue, we also find that the fraction of systems accreting at Eddington is $\sim80\%$, $\sim50\%$ and $\sim1\%$ for Pop3, Q3d and Q3nd respectively. 
However, if we consider only the subset of EMcps, these fractions increase to  $\sim100\%$, $\sim88\%$ and $\sim53\%$ because of the requirement on the detectability of the EMcp.

The lower panels of Fig.~\ref{fig:magn_athena_distro_stsi} confirm that the trade-off between sky localization and limiting flux penalises the scenario with the $\Delta \Omega = 2 \degsq$ threshold as the AGN are generally faint. 
Assuming accretion at Eddington, the number of EMcps increase, as expected. In particular, the Q3nd scenario provides slightly more EMcps than Q3d, as the luminosity depends only on the mass of the binary and not on the amount of gas available for the accretion.

The number of EMcps as a function of the magnitude of their host galaxies, observable with ELT, are shown in Fig.~\ref{fig:magn_elt_distro_stsi}. In the maximising case, most of the systems have $m_{\rm  gal,ELT}>17.5-20$ but the inclusion of obscuration and jet pushes this value up to $m_{\rm gal, ELT}>22.5$. As we move to larger apparent magnitudes, the number of fainter sources increase for all astrophysical models in a similar way. 
Most of LISA sources are hosted in faint galaxies, pushing the boundaries toward populations that are challenging to observe. Even if there is always a bright fraction of EMcps the bulk of the population is at the limit of the magnitudes currently observed.

Flux-limited samples are always dominated by faint sources, but this statement is  specifically motivated by the actual physical properties of the sources: MBHs of mass $10^3-10^7 \msun$ hosted in galaxies with mass $10^7-10^{10} \msun$ at high redshift. For sources with SNR$>10$ the median MBH mass is $10^{3.57}$, $10^{5.22}$ and $10^{5.75}$ solar masses (Pop3, Q3nd, Q3d), the median galaxy masses are 
$10^{8.24}$, $10^{7.20}$ and $10^{8.63} \msun$ (Pop3, Q3nd, Q3d) and the median redshift 5.18, 8.26 and 3.90.  In terms of absolute magnitudes the median AGN absolute magnitude is -13.01, -16.67 (Pop3, Q3d) and the median galaxy absolute magnitude is -15.95, -18.57 (Pop3, Q3d), which according to normal definition are faint sources, in line with the galaxies being dwarfs, based on their masses. Since most of the mergers are at high redshift the corresponding apparent magnitudes also are very faint, but we want to stress that the intrinsic faintness is really a property of LISA sources: small MBHs in high-z dwarf galaxies. 
Concerning Q3nd, since the Eddington ratios are incredibly small ( c.f.~Fig.~\ref{fig:Ledd_ratio_vs_mchirp}) the median absolute magnitudes of the AGN is actually a positive value. The median absolute magnitudes of the galaxies is also very very faint, -12.25, since some of the mergers occur in galaxies with a baryonic mass smaller than $10^6 \msun$.

\subsection{\label{subsec:magn_flux_distro_all_catalog} Magnitudes and fluxes distributions for the entire catalogue}

\begin{figure}
    \includegraphics[width=\columnwidth]{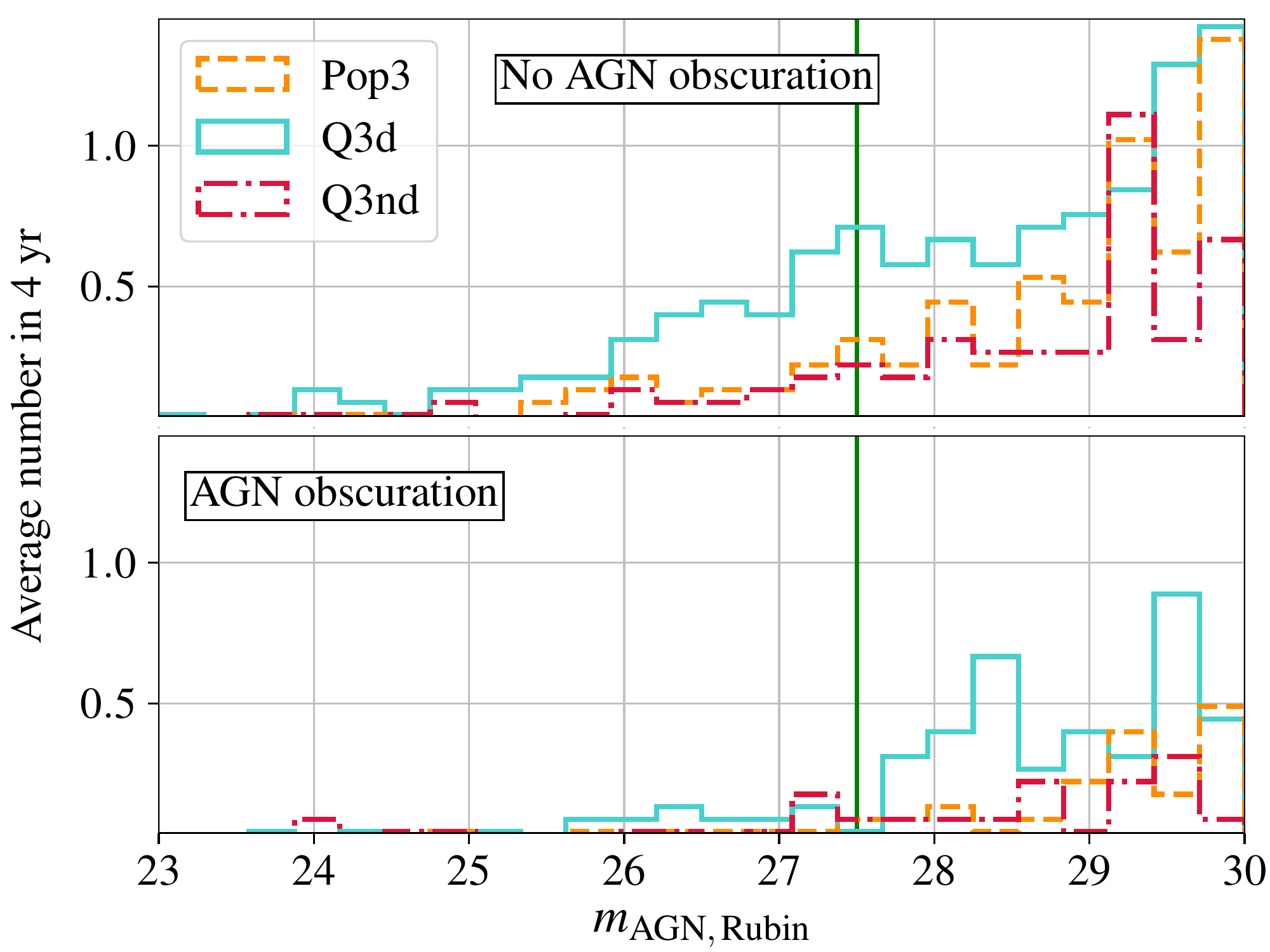}
    \caption{Same as Fig.~\ref{fig:magn_lsst_distro_stsi} but for entire catalogue. The green vertical line represents the limiting magnitude $m_{\rm AGN, Rubin, \, lim}$. }
    \label{fig:magn_lsst_distro_cp}
\end{figure}

\begin{figure}
    \includegraphics[width=\columnwidth]{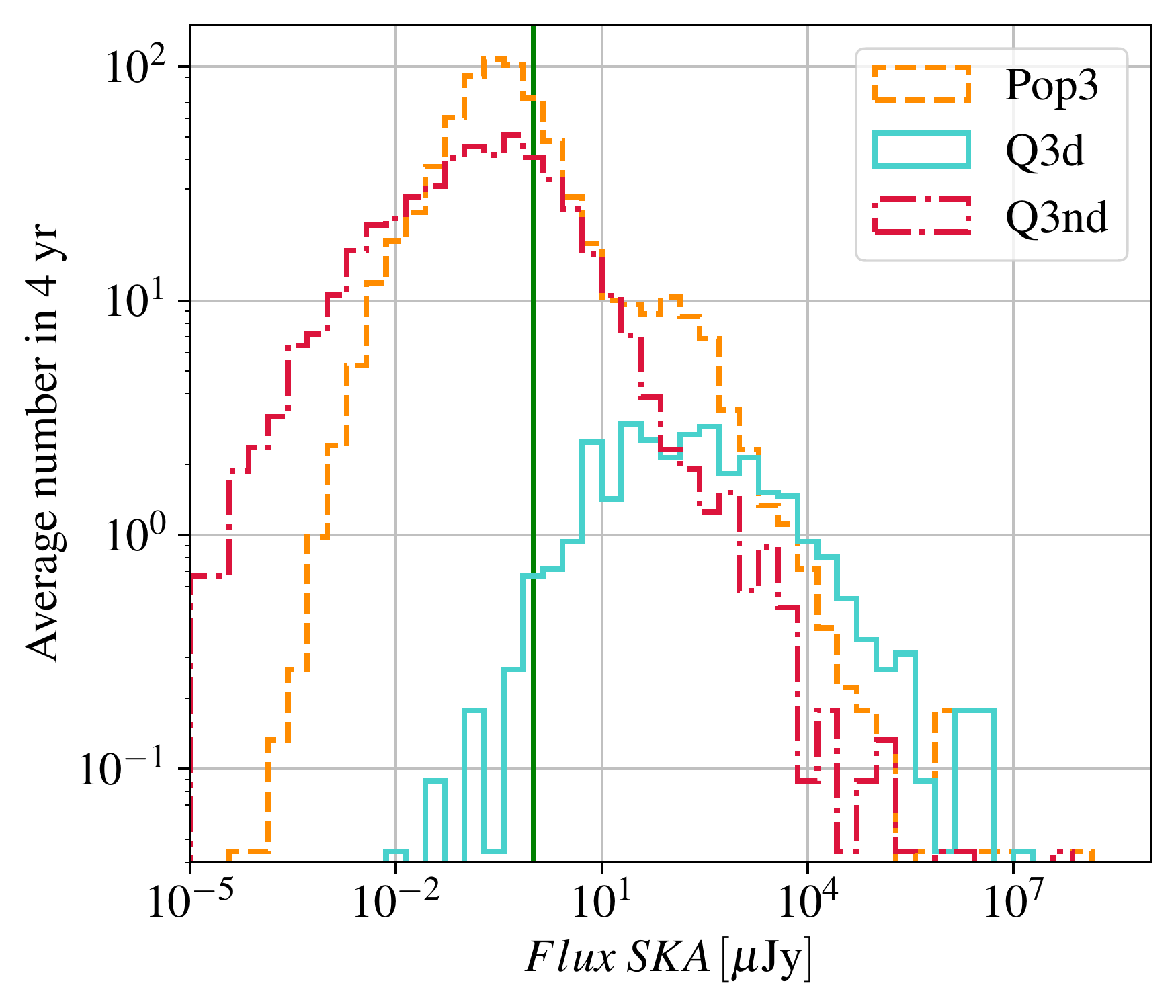} 
    \caption{Same as the upper panel of Fig.~\ref{fig:magn_ska_distro_stsi} but for the entire catalogue. The vertical green line corresponds to $1\mu \rm Jy$. }
    \label{fig:magn_ska_distro_cp}
\end{figure}

\begin{figure*}
    \includegraphics[width=\textwidth]{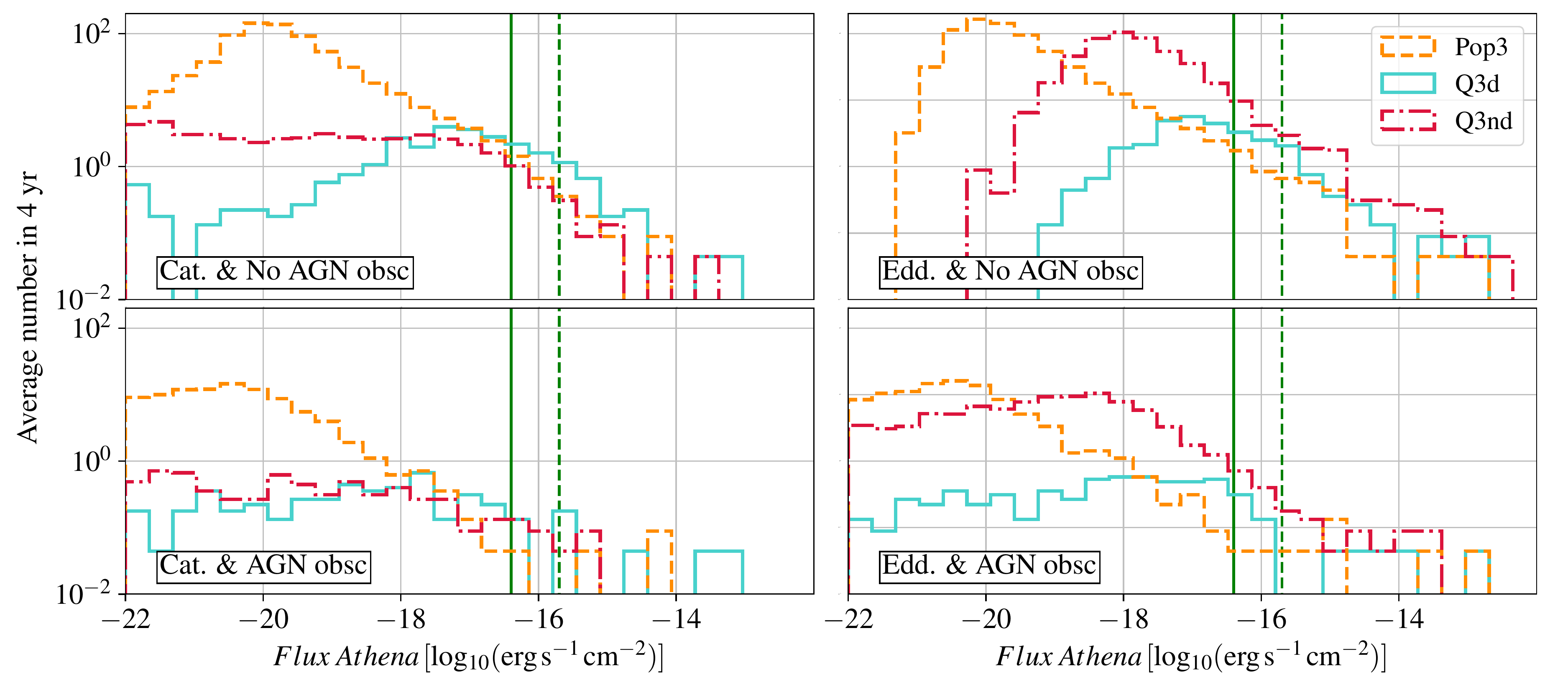} 
    \caption{Flux distributions for the entire catalogue, for the three astrophysical scenarios and different configurations. The left (right) panels correspond to the case where the accretion is computed from the catalogues (assumed at Eddington). The vertical solid 
    (dashed) green line corresponds to $F_{\rm X, \, lim} = 4 \times 10^{-17} \rm erg \, s^{-1} \, cm^{-2} $
 ($F_{\rm X, \, lim} = 2 \times 10^{-16} \rm erg \, s^{-1} \, cm^{-2} $). }
    \label{fig:magn_athena_distro_cp}
\end{figure*}

\begin{figure}
    \includegraphics[width=\columnwidth]{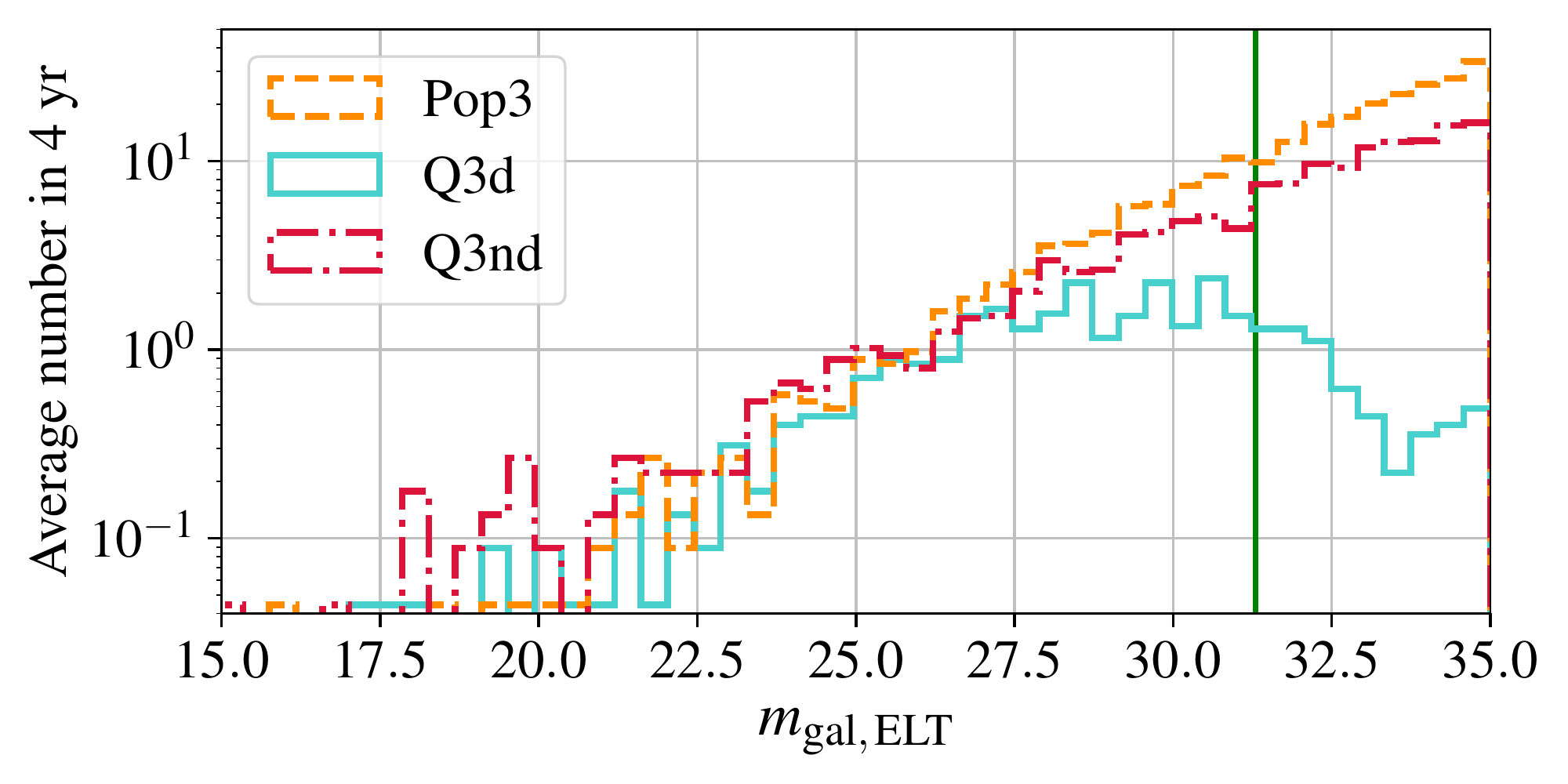} 
    \caption{ Same as Fig.~\ref{fig:magn_elt_distro_stsi} but for the entire catalogue. The vertical green line corresponds to 31.3}
    \label{fig:magn_elt_distro_cp}
\end{figure}

In this section we present the magnitude and flux distributions for the entire catalogues, i.e. without the requirement on the $\SNR$ or on the sky localization.

In Fig.~\ref{fig:magn_lsst_distro_cp} we show the magnitude distributions of the sources in our catalogues that can be potentially observed with the Rubin Observatory. Similarly to the EMcps case, the Q3d model predicts more events than Pop3 and Q3nd at all magnitudes. Moreover, the Rubin Observatory does not contribute significantly to the number of multimessenger candidates even at $m_{\rm AGN, Rubin}>27.5$ due to the intrinsically low fluxes expected from these systems. 

Moving to SKA, in Fig.~\ref{fig:magn_ska_distro_cp} we present the radio fluxes for the isotropic flare case. For all the astrophysical models, the peak at lower fluxes arises from the flare emission. For the Q3d model, most of the sources are above the detection threshold and the distribution from the entire catalogue is similar to the EMcps one (c.f.~Fig.~\ref{fig:magn_ska_distro_stsi}). 

In Fig. \ref{fig:magn_athena_distro_cp}
we show the flux distributions in soft X-ray assuming the accretion at Eddington or from the values computed in the catalogues and in the case with and without AGN obscuration.
Starting from the latter case, the X-ray fluxes for the three astrophysical models are similar above the threshold but they show different behaviour at fluxes $< 10^{-18} \rm erg \, s^{-1} \, cm^{-2} $ due to the different Eddington ratio values and BH masses. 

If we  assume Eddington accretion, the flux depends on the binary total mass: since Pop3 has the lightest binaries, it also produces the faintest sources, while Q3nd and Q3d produce brighter emission.
The inclusion of AGN obscuration leads 
to a reduction in the global number of systems with stronger tails extending at fainter fluxes. 

In Fig.~\ref{fig:magn_elt_distro_cp} we report the galaxy magnitude distributions. All  three astrophysical models are similar up to $m_{\rm gal, ELT} \sim 30$. Above this value, the Q3d model starts decreasing due to the lack of sources while Pop3 and Q3nd proceed with the same trend. Both Pop3 and Q3nd model distributions reach the peak at magnitudes that are too small to be detectable by any planned instruments so we limit the x-axis to $m_{\rm gal,ELT}=35$.

\section{\label{sec:multimodal_systems}Multi-modal systems}

\begin{figure*}
    \centering
    \subfigure{\includegraphics[width=0.33\textwidth]{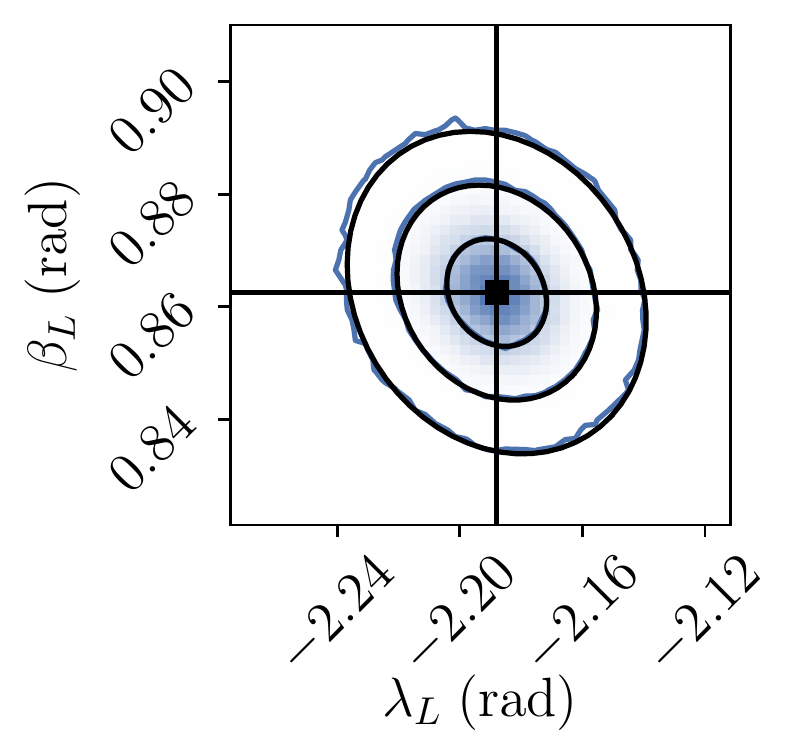}} \hspace{-0.3cm}
    \subfigure{\includegraphics[width=0.33\textwidth]{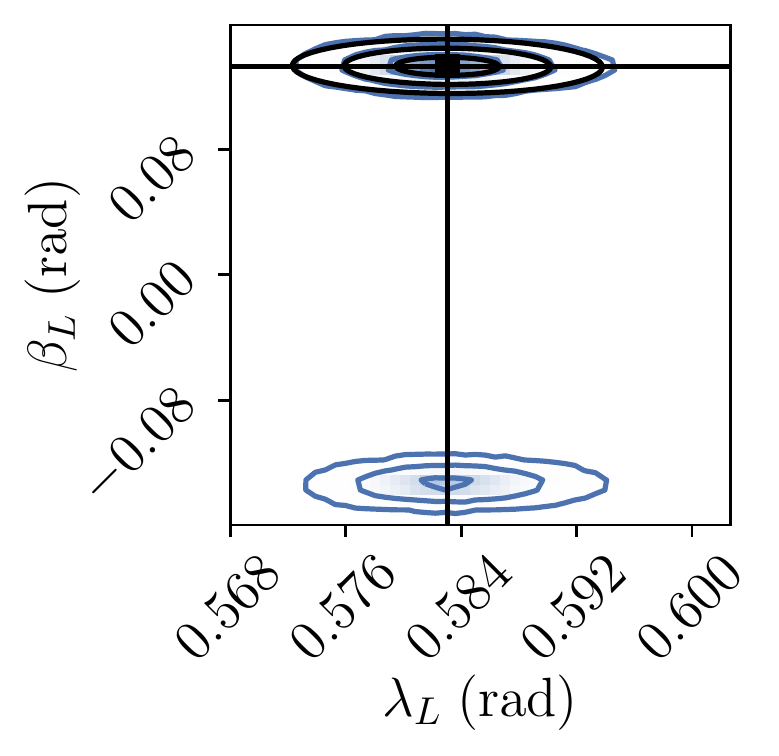}} \hspace{-0.3cm}
    \subfigure{\includegraphics[width=0.33\textwidth]{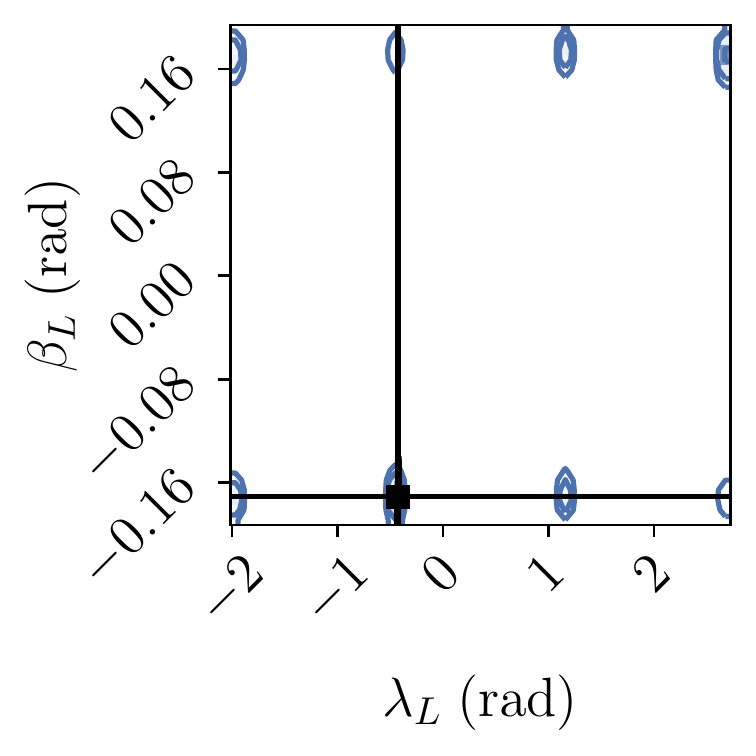}}
    \caption{Examples of multi-modal posterior distributions in the sky localization: from left to right, \emph{1mode}, \emph{2modes} and \emph{8modes} binary systems. 
    The blue contours represent the MCMC results while the black ellipses correspond to the Fisher estimates for $1\sigma$, $2\sigma$ and $3\sigma$ errors. The black square indicates the true binary position in the sky.
    The presence of \emph{2modes} and \emph{8modes} events, and the relative probability weight of the modes, can be recovered only with a Bayesian analysis.}
    \label{fig:skymodes}
\end{figure*}

LISA ability to accurately localize the gravitational waves source in the sky will strongly depend on the system's parameters, leading to a distribution of sky position uncertainties that spans several orders of magnitude \cite{Mangiagli20}. Moreover, there has been also evidence \cite{Marsat21} of systems whose sky position posterior distributions are multi-modal in the sky, i.e.~they  peak not only at the true binary position but also in other regions, symmetrically distributed in the sky. 
The emergence of these  `multi-modal events' is due to the intrinsic degeneracy in the LISA pattern functions. 
The degeneracy can be broken only if enough signal is accumulated at low frequencies, where the orbital motion of the detector provides additional information, or at high frequency, thanks to the frequency dependence of the detector response function. 

Multi-modal sky position posteriors  pose a serious challenge to the search of EM counterparts, since telescopes have to search in a larger region of the sky. 
In addition, under some conditions that we will discuss, the probability of the ``spurious'' modes is similar to the probability of the real mode (the actual binary position), further challenging the detection of a counterpart.
Fortunately, as we will see, multi-modal EMcps are relatively rare (c.f.~Tab.~\ref{tab:fraction_multimodal_events}).

In \cite{Marsat21, Sylvain_paper} it has been shown that, defining $(\beta_{T}, \lambda_{T})$ the true binary latitude and longitude, the spurious modes appear at $\beta = - \beta_{T}$ and $\lambda =  \lambda_{T} + k \pi/2$ with $k = 0,1,2,3$, for a maximum of 8 modes in the sky (one true and seven spurious).
For events with only one spurious mode, this secondary point is generally the reflected mode with $k=0$, i.e.~$(-\beta_{T}, \lambda_{T})$.
In a minority of cases (two out of the entire catalogue) it is, instead, the antipodal one, i.e.~$\beta=- \beta_{T}$ and $\lambda = \lambda_{T} + \pi $.

Among the degenerate modes, the reflected one deserves a separate discussion \footnote{In the spurious modes we expect degeneracies also in the binary inclination and polarization but these are not relevant for our scope}. 
Without going into the details (we refer the interested readers to \cite{Marsat21}), the reflected mode is exactly degenerate with respect to high-frequency effects in the response, and only LISA's motion is  expected to break the degeneracy. 
Thus, the reflected mode appears in the sky position posteriors of signals that are short enough for the LISA motion to be unimportant. 
Furthermore, the other modes also appear in the sky localization posteriors of systems that are massive enough for their waveform not to reach high frequencies.
The degeneracy leading to the other modes, in fact, are usually broken during the merger by the frequency-dependence of LISA's response function. As a consequence, the other modes also become more common in the parameter estimations performed pre-merger.

To define multi-modal events, we introduce the concept of probability for each mode, defined as the ratio between the number of samples in a mode over the total number of samples in the MCMC analysis. 
We then define as a \emph{1mode} system, a binary whose sky localization posterior has a probability larger than 5\% only in a single sky-region.  A \emph{2modes} system is such that the probability in the reflected mode is at least 5\%, and a \emph{8modes} system is such that the sum of the probability of the other six modes (the total number of modes minus the true binary position and the reflected spot) is at least 5\%.

An example of these three cases is reported in Fig.~\ref{fig:skymodes}. Unimodal events are typically well localized and the Fisher analysis provides a similar result to the Bayesian inference. 
For \emph{2modes} systems, two  spots symmetrical with respect to the equatorial plane of LISA's orbit appear and, for \emph{8modes} systems, the sky position posterior distribution presents eight different peaks located symmetrically. 
By construction, the Fisher approach, which is a local Gaussian approximation to the posterior, is not able to recover posteriors with multiple peaks. 

Some multi-modal events are potential EMcps candidates, and we must include them in our analysis.
Two key factors need to be taken into account: the sky localization area of each mode, and the corresponding mode probability.
First of all, we want to eliminate events with too wide sky localization region, because telescopes can not explore large areas in the sky. 
This cut can be performed unambiguously
for unimodal systems, but for multi-modal events  there are different approaches: the cut can be applied only to the sky localization of the primary mode, or one can choose to combine the sky area of all the modes, assuming the telescope is going to re-point to other locations. 
This choice influences the number of EMcps.
For example, if we assume a threshold of $\Delta \Omega = 10 \degsq$ and want to cut all events with larger sky localization region,
a bimodal system where the primary and secondary modes have $\Delta \Omega = 8 \degsq$ each, is an EMcp in the former approach (with a 50\% probability of missing it if the telescope does not point to the right location), but not in the latter, because the total sky area - $\Delta \Omega = 16 \degsq$ -  would be above threshold. 
Second, one can also include a requirement on the probability of the modes: for example, one could consider as viable EMcps only the events for which the probability in the primary mode is higher than 50\%; 
or one could argue that modes which probability is less than a given threshold percentage can be discarded, and the EMcp treated like an unimodal one, as far as EM telescopes are concerned.

In this work, we have decided to focus only on the sky localization of the primary mode (i.e. the sky-region where the binary actually stands)  as the criterion to select viable EMcps, and  to apply no requirement on the probability of the other modes. 
In other words, the number of EMcps is given by the systems with detectable EM counterpart, and a sky localization region below threshold in the primary mode. 
This simplification is possible because, as we will show below, events with multi-modal sky posteriors are a minority of all cases, and furthermore, for most bimodal posteriors there is a clear hierarchy in the probability of  the primary and the secondary mode. 
Therefore, the final number of EMcps does not depend excessively on the selection criterion.
Note that, both eliminating from the catalogues all the events without detectable EM counterpart, and considering the sky localization region emerging only from the post-merger parameter estimation analysis, help in reducing the number of multi-modal events: in fact, multi-modal posterior distributions in the sky localization are more frequent at high redshift and for parameter estimation analyses performed pre-merger, when the signals are shorter and have lower SNR~\cite{Sylvain_paper}.

\begin{figure*}
    \includegraphics[width=1.\textwidth]{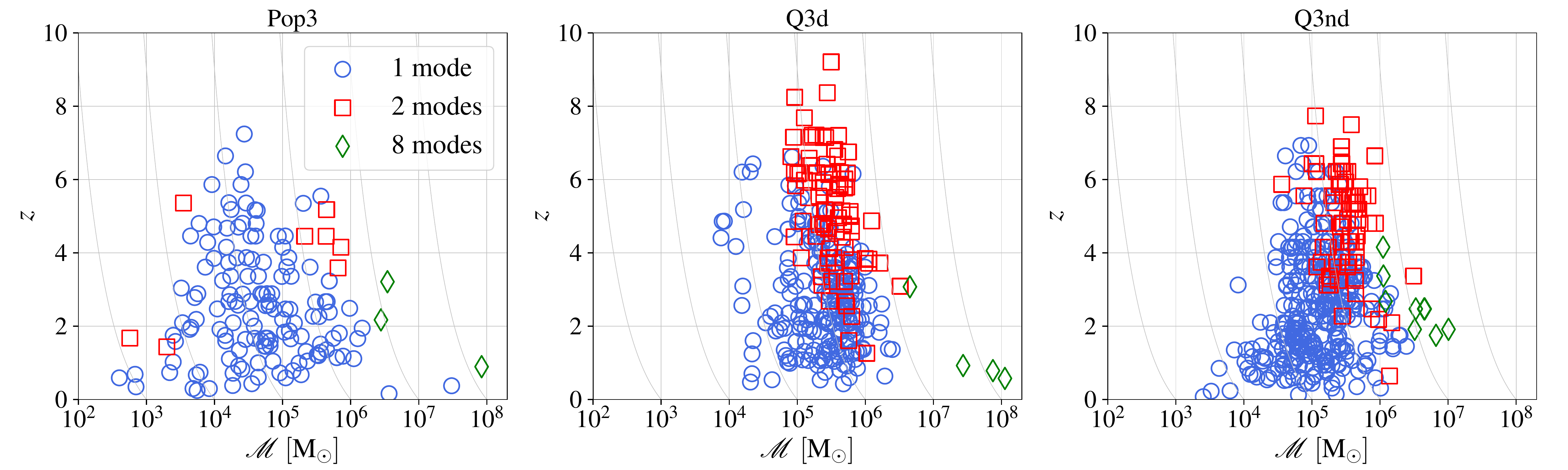} 
    \caption{Distribution of the \emph{1mode}, \emph{2modes} and \emph{8modes} EMcps in the $z-\mchirp$ plane in the maximising case for the three astrophysical scenarios. The grey solid curved lines in background correspond to constant redshifted chirp mass values.}
    \label{fig:multimodes_scatter}
\end{figure*}

\begin{figure}
    \includegraphics[width=0.5\textwidth]{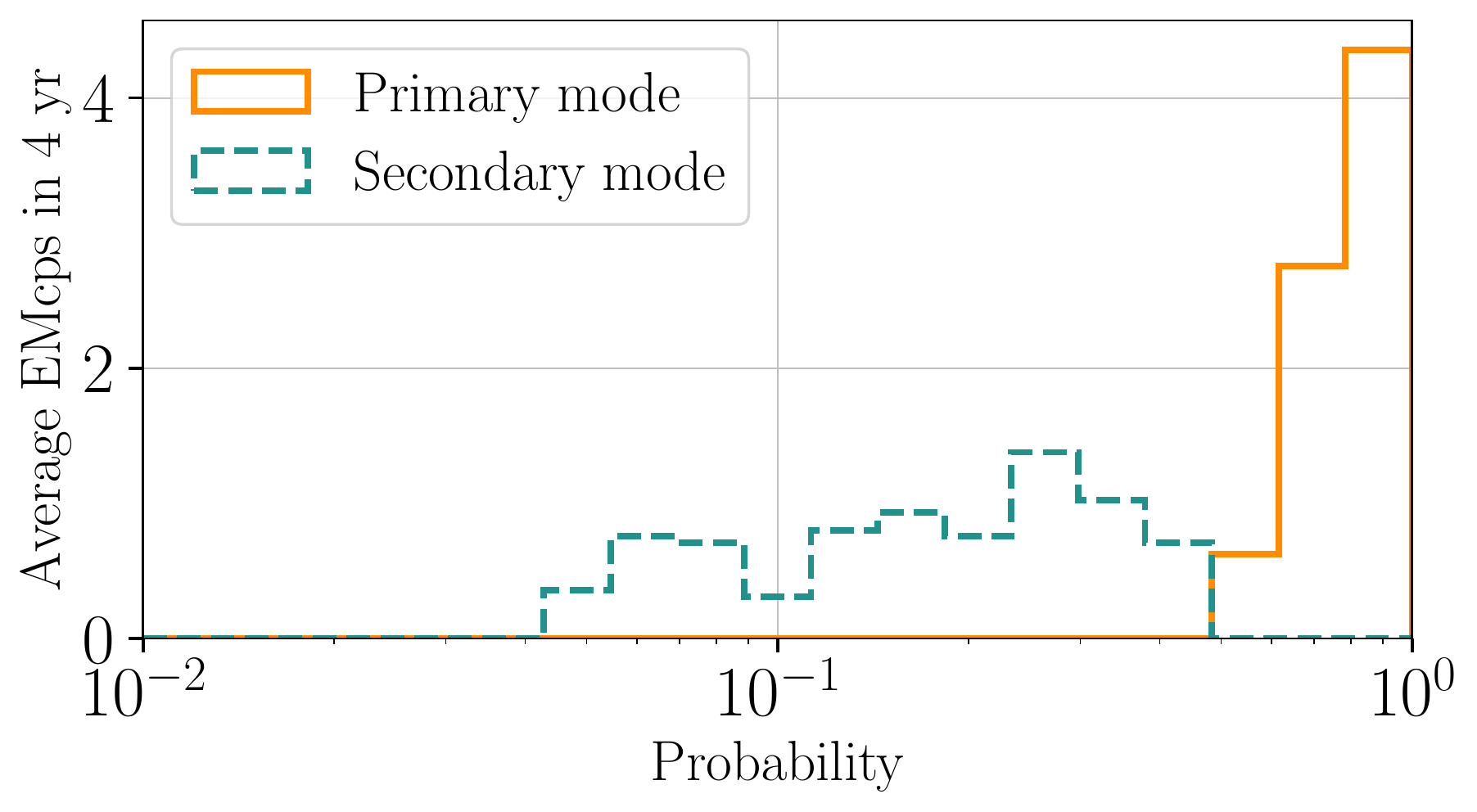} 
    \caption{Number of \emph{2modes} EMcps as a function of the probability of the sky position modes, in the maximising scenario, and combining all astrophysical models.
    The solid (dashed) line corresponds to the probability of the primary (secondary) mode. 
    Note that, by construction, the same EMcp appears twice in this figure.
    For \emph{2modes} systems, the primary mode is always more probable than the secondary.}
    \label{fig:2modes_distro}
\end{figure}

\begin{figure}
    \includegraphics[width=0.5\textwidth]{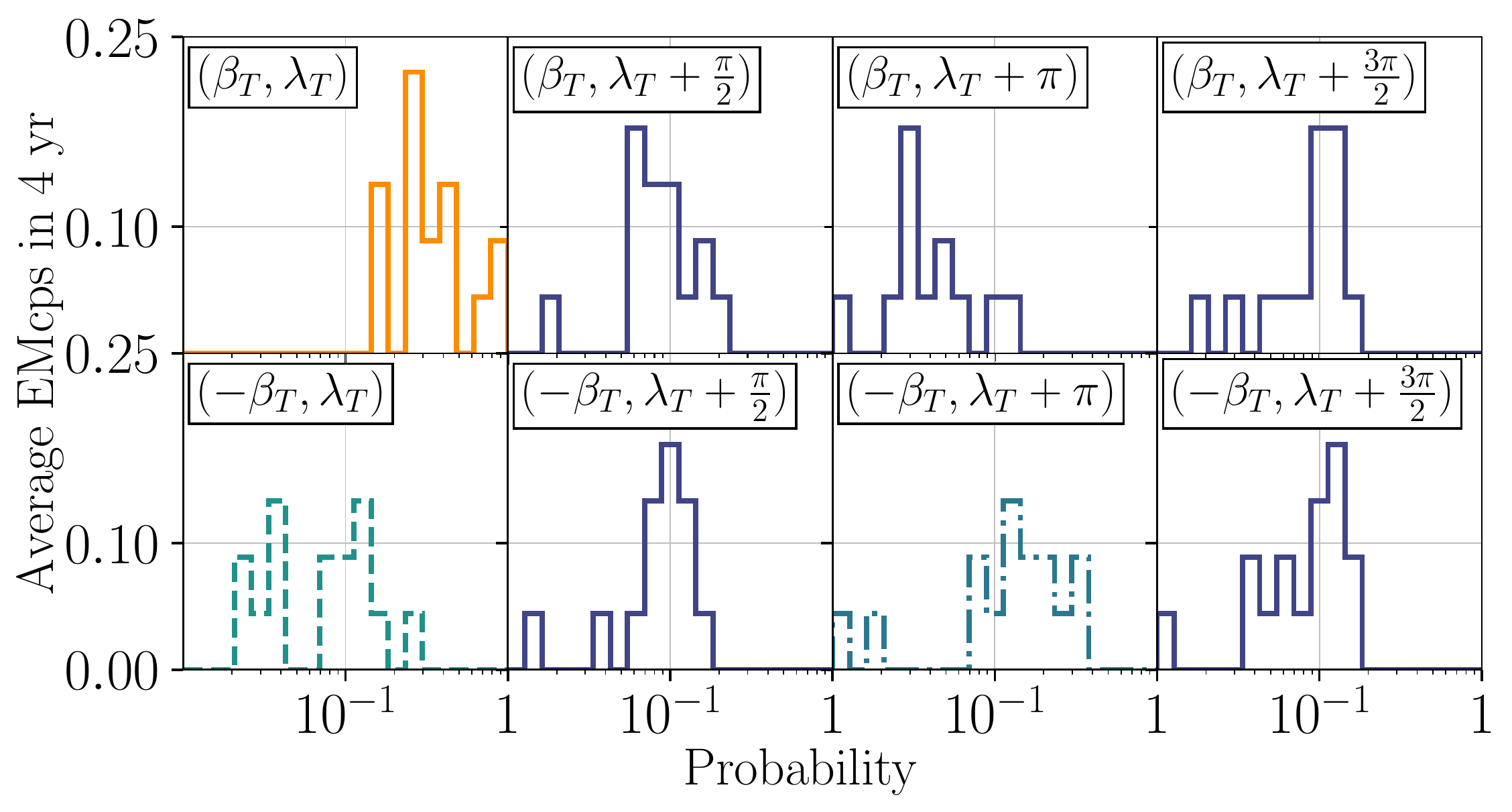} 
    \caption{Number of \emph{8modes} EMcps as a function of the probability of the 8 sky location modes, in the maximising scenario, and combining all the astrophysical models.
    Each panel shows a different spot in the sky, according to the legend. 
    The solid orange line in the leftmost panel, upper row, corresponds to the 
    real sky position. 
    The dashed and dotted-dashed red lines correspond to the reflected and antipodal sky positions respectively, while the teal solid lines correspond to the other modes. 
    Note that, by construction, the same EMcp appears 8 times in this figure.}
    \label{fig:8modes_distro}
\end{figure}

\begin{table}
\caption{\label{tab:fraction_multimodal_events} 
Number of \emph{1mode}, \emph{2modes} and \emph{8modes} EMcps in 4 yrs of LISA observation, in the maximising case, and for the three astrophysical models.
The sum of all events in each astrophysical model corresponds to what given in Table \ref{tab:stsi_combined_strategies}. }
\begin{ruledtabular}
\begin{tabular}{c | m{1.4cm} | m{1.4cm} | m{1.4cm}  } 
   &  \emph{1mode} & \emph{2modes} & \emph{8modes}   \\
\Xhline{0.1pt}
Pop3   & 6.0  & 0.31  & 0.13    \\
Q3d  & 10.7 & 3.9 & 0.18  \\
Q3nd  & 16.8 & 3.5 & 0.4  \\
\end{tabular}
\end{ruledtabular}

\end{table}

In Tab.~\ref{tab:fraction_multimodal_events} we report the fraction of \emph{1mode}, \emph{2modes} and \emph{8modes} EMcps in the maximising case (the minimising scenario provides similar results, just rescaled). 
For all astrophysical models, the largest fraction of EMcps is unimodal, while bimodal EMcps contribute  5\%, 26\% and 17\% of the total rates for Pop3, Q3d and Q3nd respectively. 
By contrast, \emph{8modes} events constitute less than 0.5 EMcps in 4 yrs of LISA observation.

Fig.~\ref{fig:multimodes_scatter} shows the distribution of \emph{1mode}, \emph{2modes} and \emph{8modes} EMcps, in the maximising case, in the $z-\mchirp$ plane. 
It appears clearly that the vast majority of the EMcps population is constituted of unimodal events. 
Furthermore, the redshifted chirp mass $\mchirp_z$, being the quantity that enters the waveform,
influences the appearance of multi-modal posteriors. \emph{2modes} events 
tend to have masses such that $\mchirp_z \gtrsim 10^6 M_\odot$, and relatively high redshift, $z \gtrsim 2$.
Events with high chirp mass and redshift, in fact, have GW signals long enough in the LISA band for the high frequency instrument response to play a significant role in breaking the pattern function degeneracy.
However, most of the signal is accumulated in the last days, so the effect of LISA motion is insufficient to eliminate the reflected sky position, which remains degenerate.
At lower chirp mass and redshift, on the other hand, the combination of the high-frequency response with the motion of the detector fully eliminates the degeneracy and the events are unimodal. 
However, even if it is possible to identify a general trend, the two sub-populations of \emph{1mode} and \emph{2modes} events do overlap in the $z-\mchirp$ plane,
because redshift and $\mchirp_z$ are not the only quantities affecting the parameter estimation, and there is a large dispersion according to the orientation angles. 
Regardless of the astrophysical model, the \emph{8modes} systems are high chirp mass MBHBs, for which LISA will be able to observe only the merger and ringdown, gathering little information from the constellation orbital motion; furthermore, their GW signal will not reach high enough frequencies for the frequency-dependence of the detector response to help.

Although \emph{2modes} systems seem to constitute a significant portion of the total EMcps, especially for the massive astrophysical models Q3d and Q3nd, this is partly caused by the fact that, to identify an event as bimodal, we impose a relatively low threshold to the probability of the secondary mode, i.e.~5\%.
In this regard, it is instructive to look at the probability weight of each mode. 
In Fig.~\ref{fig:2modes_distro} we present the number of bimodal EMcps as a function of the  probability in the primary and secondary modes, for all astrophysical models. 
It is clear that the primary mode is always more probable than the secondary one, which mitigates the risk, for a substantial fraction of the EMcps, of missing the counterpart if telescopes are pointed only at the primary mode. 

In Fig.~\ref{fig:8modes_distro} we show the number of \emph{8modes} EMcps in each octant of the sky, as a function of the probability of the sky position mode. 
While the primary mode remains always more probable, the seven spurious modes are rather equiprobable, with probability that can be as large as 10\%. 
This is likely to constitute a serious issue for the search of the EM counterparts of \emph{8modes} events. 
Fortunately, our results show that \emph{8modes} events are not going to contribute significantly to the total number of EMcps after merger (c.f.~Table \ref{tab:fraction_multimodal_events}).

Following these results, 
in this work we have decided to base the selection of multimessenger candidates that can become EMcps only on the requirement that the sky localization region of the primary mode after merger is small enough, as previously described.
We decided not to apply, to the so obtained number of EMcps, a
correction factor representing the fraction of events that might be missed on average because telescopes are mistakenly pointed at the wrong spot in the sky.

\section{\label{sec:discu_and_concl}Discussion and conclusions}
MBHBs are key sources for multimessenger astrophysics with LISA, as they are expected to merge at the center of galaxies, where the accreting gas might produce an EM counterpart to the GW signal. 
In this paper we presented the number and characteristics of both multimessenger candidates (i.e.~MBHB mergers with a detectable EM emission) and of EMcps (i.e. multimessenger candidates which can be localised well enough via their GW emission). We analysed different astrophysical scenarios for the MBHBs formation, and modelled the EM emission by combining some selected, current and future, EM facilities.

We took as input the results of the SAM code developed in \cite{Barausse12, Sesana14_spin_evolution,Antonini15_1, Antonini15_2} to infer the mass and redshift distributions of the merging MBHBs, as well as the properties of their host galaxies. 
For each MBHB event, we first computed the signal-to-noise ratio in LISA, to assess the number and distribution of the events detectable purely from the GW side. 
We then exploited the MBH and host galaxy properties to compute the expected EM emission at several wavelengths, from radio to soft X-ray. 
We considered three observational scenarios: EMcp identification and redshift determination both with the Vera Rubin Observatory; EMcp identification with the SKA and redshift determination with the ELT; EMcp identification with Athena and redshift determination again with the ELT. 
These observational scenarios cover the entire EM spectrum; naturally, the same system can be detectable by one or more observatories at the same time.

For the subset of events classified as multimessenger candidates, we estimated the errors on the binary parameters, especially the sky localization, 
inferred by the GW signal with LISA, performing Bayesian parameter estimation as in \citet{Marsat21}. 
We then selected as EMcps only the systems for which the sky localization is smaller than given thresholds, appropriately chosen following the capacities of the EM telescopes. EMcps are therefore those MBHB mergers that can both be exploited for subsequent astrophysical studies, and be used as standard sirens for cosmology, since one can infer their luminosity distance from the GW signal, and their redshift from the EM counterpart.

We focused especially on two scenarios of viable EMcps: one maximising their number, in which we assumed that the radio flare emission is isotropic and that there is no AGN obscuration; and one minimising their number, accounting for beamed radio emission and AGN obscuration. 
In the maximising scenario we predict  $14.9$ ($6.8$) \{$20.9$\} EMcps for the Q3d (Pop3) \{Q3nd\} astrophysical models respectively,  in 4 yrs of LISA observations. 
In the minimising scenario, these rates decrease to 
$3.4$ ($1.7$) \{$3.4$\}, respectively.
The collimation of the radio jet and the AGN obscuration by hydrogen and dust are the two features most affecting our results as far as the detection of the EM emission is concerned. 

Removing the requirement of the redshift determination does not change significantly the number of MMcands and EMcps observed with the Rubin Observatory and Athena: the cuts in magnitude imposed by the detection with these facilities already select sources at relatively low redshift. Concerning the SKA, on the other hand, we find that the number of MMcands increases by a factor $\sim 2.5-5$ when removing the redshift identification requirement, because SKA can reach higher redshifts than ELT. 
Note that the number of EMcps always remains the same: these systems have, in fact, low redshift due to the requirement on the sky localisation.

We find that EMcps can be detected up to $z\sim 6-8$ and they have typically $\mchirp \sim 10^4 - 10^6 \msun$, because these systems have a sufficient amount of gas available for accretion and they are localized with enough accuracy by LISA. 
Considering each observational scenario separately, we find that the Rubin Observatory is the instrument providing the smallest number of EMcps, with on average less than one event in four years of joint observation with LISA.
On the other hand, SKA+ELT provide the largest number of EMcps, if the radio flare emission is isotropic; however, the EMcps rate is drastically reduced if the radio flare and jet emissions are collimated in a jet cone with opening angle  $ \theta \sim 30^\circ$, and goes to zero for $ \theta \sim 6^\circ$.
Therefore, under the plausible assumption of a collimated radio emission, only the presence of Athena in conjunction with LISA would re-enable the possibility of having at least a few EMcps in 4 yrs, assuming, though, that the X-ray emission is not affected by dust obscuration. Introducing
AGN obscuration leads in fact to a significant drop in the EMcp rates. 
Interestingly, restricting the sky area to the size of the Athena FOV, thereby reaching deeper fluxes, 
increases the number of EMcps, rather than
scanning a larger area of the sky with higher flux limit. 
We therefore identify the first observational strategy as the one capable of maximising the opportunity of joint LISA-Athena observations.

The number of multimessenger candidates changes in the same way as the one of EMcps, when changing the observational strategy. 
We found that the Pop3 astrophysical model leads to more multimessenger candidates than the massive models Q3d and Q3nd, but most of them do not satisfy the sky localization requirements, because they are too light and at high redshift. Consequently, these events cannot be classified as EMcps, so that the final number of EMcps in the Pop3 astrophysical model is overall smaller than in the massive models.
It is also important to remark that the vast majority of the MBHB mergers in the SAM catalogues are characterised by an EM emission much fainter than the threshold magnitudes and fluxes of the EM observatories considered. Since these facilities are representative of the planned future telescopes, this shows, to the best of our present knowledge, how challenging multimessenger MBHB  observations with LISA will be.

We also found that a fraction of multimessenger candidates present multi-modal sky posterior distributions, characterised by two or eight spots in the sky, symmetrically distributed over the sphere. 
We promoted these events to EMcps when their GW-inferred sky localization is smaller than the selected thresholds in the primary mode only, neglecting the other modes. 
We found that bimodal events can contribute up to $\sim 25\%$ of all EMcps; however, the posterior probability of the primary mode is always larger than the one of the secondary one. 
\emph{8modes} events, on the other hand, have nearly equiprobable sky-localisation modes, but they contribute less than 0.5 EMcps in 4 yrs of LISA observation, so they play a minor role and do not affect our results. 

The present analysis has also some caveats. 
Concerning the catalogues, our results are based on the same MBH physics described in T16. 
More recently, new predictions for the MBHB merger rates have been published \cite{Barausse20}. In this work, the authors refined the modeling of the time delays,  accounting for the baryonic components at galactic merging scale and implementing better prescription for dynamical friction at smaller scales ($<100$ pc). They also included the effect of supernova feedback that may reduce the amount of available gas in low-mass galaxies.
We compared the two catalogues and found no significant differences in the number of both intrinsic and detected i.e.~(${\rm SNR}>10$) MBHB mergers for the Q3d model (the `heavy seed' model in the recent work). 
We note especially that, within the Q3d model, the redshift distributions provided by the new analysis are skewed toward lower redshift, so we expect to recover a similar, or even larger, number of EMcps.
On the contrary, the new results predict more (less) MBHB mergers for the Q3nd (Pop3) astrophysical models. 
Therefore, the results obtained in this work might be overestimated for Q3nd and underestimated for Pop3. 

Concerning the modeling of the EM emission and of its detection, several considerations are in order.
The effective number of EMcps might be reduced by  difficulties linked to the follow-up strategy that we have not addressed, like, for example, how to deal with the situation of two, or more, coincident MBHB mergers in LISA or the presence of other sources with time-varying EM emission in the sky localization region provided by LISA. Concerning the last point, \citet{Lops_paper} explored the possibility to identify the host galaxy of MBHBs mergers starting from a simulated portion of the Universe. Their work can be considered complementary to ours since they focused more extensively on the exploration of galactic fields around MBHBs mergers.
Similarly we did not take into account the possibility of a time delay of weeks to years between the GW signal and the peak of the EM emission. Such long delays would seriously impact the possibility to identify unambiguously the host galaxy since searching for a transient in a large patch of the sky with deep ToOs is challenging. However, a possible solution might be to look for modulated emission in archival data once a transient is detected.
Also, we used the post-merger localization of the GW signal by LISA,
which is drastically better than the pre-merger one, but at the same time we fully neglected the possibility that some EMcps might be detected already before merger.
In order to evaluate the bolometric luminosity in the Eddington emission case, we used the mass of the binary, while for post merger emission it would have been more appropriate to use the mass of the remnant black hole, since the GW emission is expected to subtract energy and angular momentum from the system. 
However, we found that this leads to no significant difference in our estimates, as the binary and remnant BH masses are very similar \cite{2012ApJ...758...63B}.

We also chose to be as optimistic as possible and did not apply any correction factor to account for the actual sky coverage of telescopes such as SKA or the Rubin Observatory, which will observe at best half the sky. 
In the worst case scenario, this would lead to the loss of half of the EMcps, due to Earth orientation.

Regarding Athena, our results suggests that observing a single tile (for example, the one with the highest posterior probability resulting from the LISA parameter estimation), might be a successful strategy to detect the X-ray counterpart. 
However, this result needs confirmation via a realistic end to end simulation of the observing strategy, which is behind the scope of the current work. 
For example, each tile and the corresponding observing time might be weighted with the posterior probability from the GW measurement.

Regarding the LISA data analysis, we did not account for data gaps in the detector output. 
This might be justified, as at least the gaps expected in connection with the standard LISA maintenance can be avoided by triggering a protected period around the predicted MBHB merger time, if the binary is detected sufficiently early before the merger. 
However, data gaps might still be problematic for massive systems entering in LISA band already in the merger or ringdown phase, since there will be no time to postpone a scheduled maintenance.
Moreover in this work, we perform the parameter estimation only for the MBHBs with detectable EM emission. In the future, we plan to perform the parameter estimation of all the binaries present in the SAM catalogues, independently of whether a detectable EM emission is associated to them or not \cite{Sylvain_paper}, to have a more comprehensive view.

In this work, we restrict ourselves to MBHBs with circularized orbits and spins aligned to the orbital angular momentum, thus neglecting the possibility of misaligned spins inducing precession, as well as the possible presence of eccentricity. We leave the investigation of the localization of sources in the presence of these effects for future work.

Finally, the actual distribution of observed MBHBs might arise from a combination of the three  astrophysical models we adopted in this study (see for example \cite{Volonteri10_mixture_seed, Spinoso22}). For simplicity, here we did not considered mixed astrophysical formation models.

\begin{acknowledgments}
We wish to thank Stanislav Babak, Jonathon Baird, Enrico Barausse, Robert Caldwell, Monica Colpi, Andrea Comastri, Giancarlo Ghirlanda, Matteo Guainazzi, Clotilde Laigle,  Henry Joy McCracken, Antoine Petiteau, Alberto Sesana, Lorenzo Speri, Alexandre Toubiana and Maxime Trebitsch for useful comments and discussions. 
AM and CC acknowledge support from the postdoctoral fellowships of IN2P3 (CNRS). SM, NT and HI acknowledge support from the French space agency CNES in the framework of LISA.
Numerical computations were performed on the DANTE platform, APC, France.
\end{acknowledgments}




\bibliography{bibliography.bib}

\clearpage
\appendix
\section{\label{sec:app_to_compare_with_Tamanini16} Comparison with previous results}

In this appendix we compare our results to previous works in the literature. We start comparing our estimates with \citet{Belgacem19} (hereafter `B19') where the authors adopted the same catalogues but with different LISA sensitivity,  GW waveform and EM counterpart modeling. We have seen that the SKA+ELT configuration provides the largest number of EMcps. Therefore, to perform a closer comparison between results,   we assume that the total luminosity is always $L_{\rm radio} =  L_{\rm flare}+L_{\rm jet}$, i.e. we assume an isotropic jet, as it was computed in the original works.

Assuming 4 yr of LISA observations, they predict 208.6, 30.5 and 471 MBHBs for Pop3, Q3d and Q3nd, respectively. These values are not reported in the original paper but the authors shared the set of data with us. These values can be compared with the values reported in Tab.~\ref{tab:total_event_detected}.
For the massive models the values are similar because these sources are typically detected with large $\SNR$ therefore different waveforms or sensitivity curves do not affect significantly the number of detected events. 

However, they predict more detections in the Pop3 scenario. In order to understand this discrepancy we focus on: 1) the waveform and 2) LISA sensitivity. In B19 the authors adopted the PhenomC \cite{Santamaria10} to compute the $\SNR$ while we adopted the PhenomHM.
The latter includes the contribution of higher harmonics in the GW signal, which are relevant at merger and ringdown. However Pop3 MBHBs are ``light'' systems so they merge at high frequency in LISA and we expect that the contribution from higher harmonics is negligible. 

For the sensitivity curve they adopt the so-called ``ESACall v1.1'' \cite{esacall} which performs better at high frequency with respect to the current ``SciRDv1''. Since the high frequency sensitivity impacts the detectability of light systems, a better sensitivity at those frequencies is expected to increase the number of detected events and might explain the discrepancy.

We can compare the number of EMcps in B19 with our results . This comparison is not trivial since the number of EMcps depends on the parameter estimation so we limit to report the general differences and propose qualitative explanations to address the different expected  numbers of EMcps. 

Even if we remove the Athena+ELT case, we predict 6.8, 14.9 and 20.9 EMcps, while in B19 they predict 13.6, 14.7 and 28.3 EMcps for Pop3, Q3d and Q3nd, respectively. 

For Pop3 we predict $\sim50\%$ less EMcps with respect to B19 due to the lower number of multimessenger candidates and to the more realistic parameter estimation. Note that the reduction probably is not due to the reduced number of detected events as these systems would be detected with low $\SNR$ so the sky localization would not be likely below the thresholds.
For Q3nd we predict ~$30\%$ fewer EMcps, compatible with the loss in the number of multimessenger candidates, while for Q3d the predictions are similar since these mergers happen at lower redshift.

In principle we can also perform the same comparison with the results in T16. However, in that work, the authors explored LISA ability to constrain cosmological parameters under different LISA configurations. Nowadays those configurations are out-of-date so a direct comparison is non trivial. The configuration closer to the SciRDv1 is the one labeled as `N2A2M5L6'. Implementing the old LISA curve and redoing the parameter estimation is extremely expensive, so we limit to point out that the old LISA sensitivity performs better than the current one at high frequency, leading to a significant increase in the number of EMcps, especially for Pop3.

\section{\label{sec:app_parameter_estimation} GW analysis}

\begin{figure*}
    \subfigure{\includegraphics[width=1\textwidth]{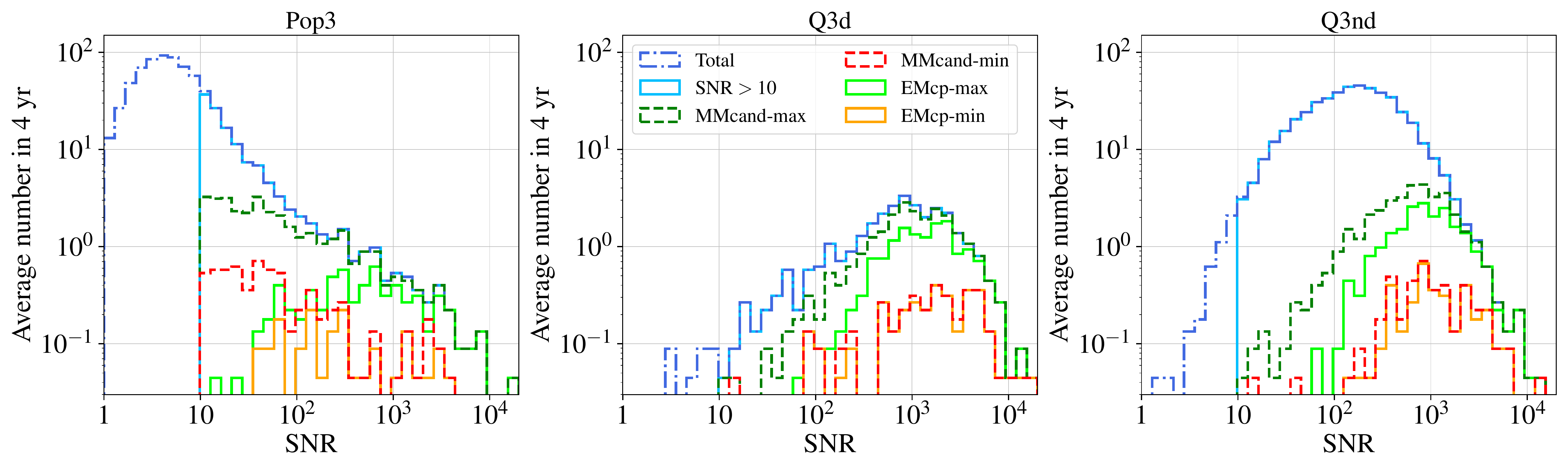}} \\ 
    \vspace{-0.4cm}
    \subfigure{\includegraphics[width=1\textwidth]{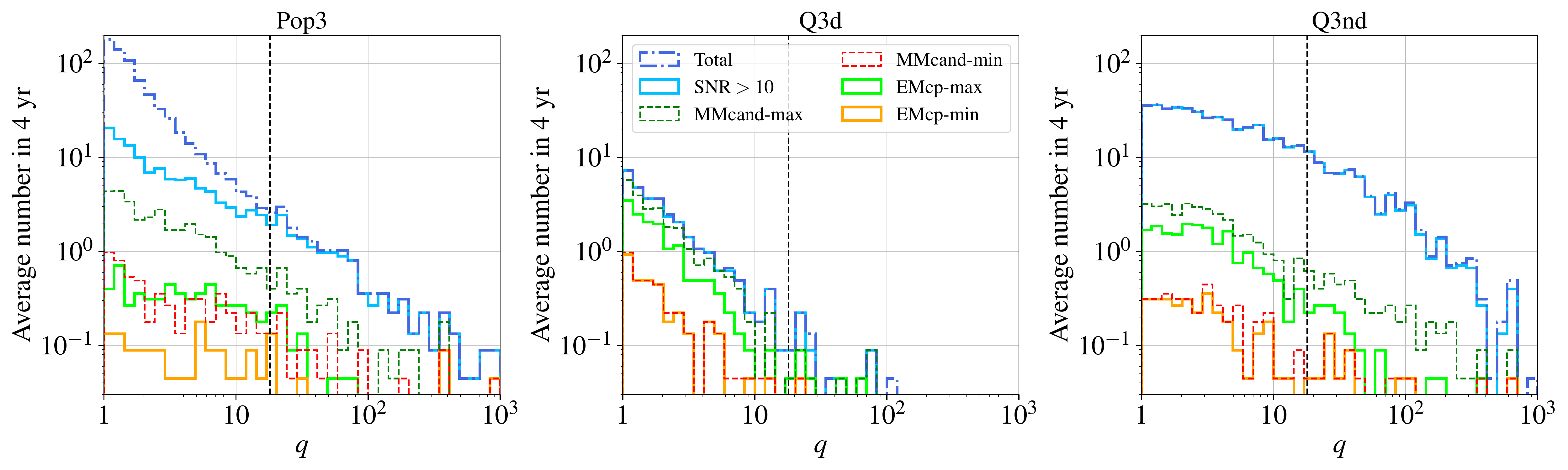}} 
    \caption{Same as  Fig.~\ref{fig:z_distro_stsi}  for the $\SNR$ and $q$. In the lower panels, the black dashed line is at $q=18$. The requirements on the sky localization and detectability of the EMcps select the systems with largest $\SNR$ and mass ratio close to unity.}
    \label{fig:snr_distro_cat}
\end{figure*}

\begin{table}
\caption{\label{tab:fraction_stsi_q_smaller_20} 
Numbers  of EMcps in 4 yrs considering only systems with $q<18$. In parenthesis, we report the rates from Tab.~\ref{tab:stsi_combined_strategies} to ease the comparison. 
}
\begin{ruledtabular}
\begin{tabular}{c | m{2.4cm} | m{2.4cm}    } 
   &  Maximising & Minimising    \\
\Xhline{0.1pt}
Pop3   & 5.5 (6.4) & 1.3  (1.7)  \\
Q3d  &  14.58 (14.8) &  3.3 (3.4)  \\
Q3nd  &  19.2 (20.7) &  3.0 (3.4) \\
\end{tabular}
\end{ruledtabular}

\end{table}

In this appendix we focus on the GW analysis of our systems. In Fig. ~\ref{fig:snr_distro_cat} we present the average number of multimessenger candidates and EMcps as function of the $\SNR$ and mass ratio $q$. As a general trend, highest $\SNR$ values lead to a better parameter estimation, so the sky localization requirement selects naturally the systems with largest $\SNR$ as EMcps. 
In Pop3 model we have multimessenger candidates even at small $\SNR$, however their parameter estimation is not good enough to promote them at the rank of EMcps while in the massive models the multimessenger candidates have always $\SNR>10$. Therefore there might be a population of low mass systems that will not be detected by LISA but that might be suitable for multimessenger studies and be accessible with Einstein Telescope \cite{et} or future deci-hertz \cite{decigo} detectors.

About the mass ratio, the multimessenger candidates and EMcps distributions for the Q3d and Q3nd models are similar to the total distribution from the catalogues. However, for Pop3, the detectability of the EM counterpart selects also systems with $q>100$: these binaries typically host a massive BH, which explains the large mass ratio, and they are at small redshift because such massive BHs require time to form in Pop3 model.

From our distribution it is evident that there is a subset of MBHBs with very large mass ratio, up to $q\sim 10^2-10^3$. As described in Sec.~\ref{sec:GW_signal}, we adopted the PhenomHM for the GW analysis. This waveform is calibrated with numerical relativity simulations performed up to mass ratio $1:18$
so the signals for $q>18$ are based on an extrapolation of the current results and might not be representative of the estimates
that an accurate waveform would produce in this range. For this reason, in Tab.~\ref{tab:fraction_stsi_q_smaller_20} we report the number of EMcps in the maximising and minimising models if we consider only the events with $q<18$.
For Q3d the numbers remain unchanged due to the fact that the vast majority of the systems has $q<18$, however the approximation in the waveform might affect the parameter estimation of $\sim20\%$ ($\sim10\%$) of the cases in Pop3 (Q3nd) astrophysical model.
Finally the PhenomHM is validated for spin magnitudes up to $0.98$ for equal mass binaries. Even if in our simulations we have MBHBs with $\chi \sim 0.99$, they are a minority and we do not expect significant differences from the parameter estimation.

\section{\label{sec:app_plot_for_discussion} Useful figures for discussion}
\label{app:useful_plot}
In this appendix, we limit to report some figures that are interesting for the discussion in Sec.~\ref{sec:results}.

\begin{figure*}
    \includegraphics[width=\textwidth]{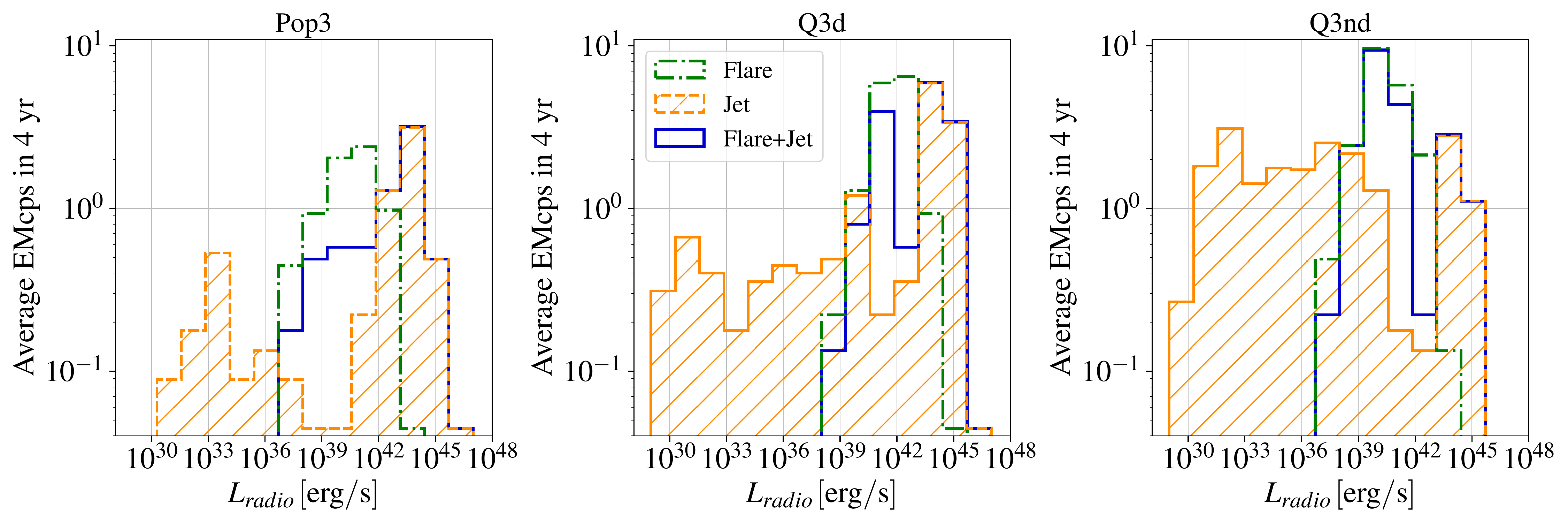} 
    \caption{Average number of EMcps as function of the luminosity of the radio emissions without imposing that the observer must be on axis. We show the EMcps emission distributions of the jet, flare and of their sum, detected with SKA+ELT for the three astrophysical scenarios, as clarified by the legend.  At high luminosity, the total emission overlaps completely with the jet luminosity, which dominates the bright end of the distribution. The presence of a floor value of $\epsilon_{\rm Edd}= 0.02$ in the flare expression (Eq.~\ref{eq:Lflare}) further modulates the flare luminosity, avoiding $L_{\rm flare}<10^{36} \rm erg/s$ values.
    }
    \label{fig:Lradio_fluxes_distribution}
\end{figure*}

\begin{figure*}
    \includegraphics[width=\textwidth]{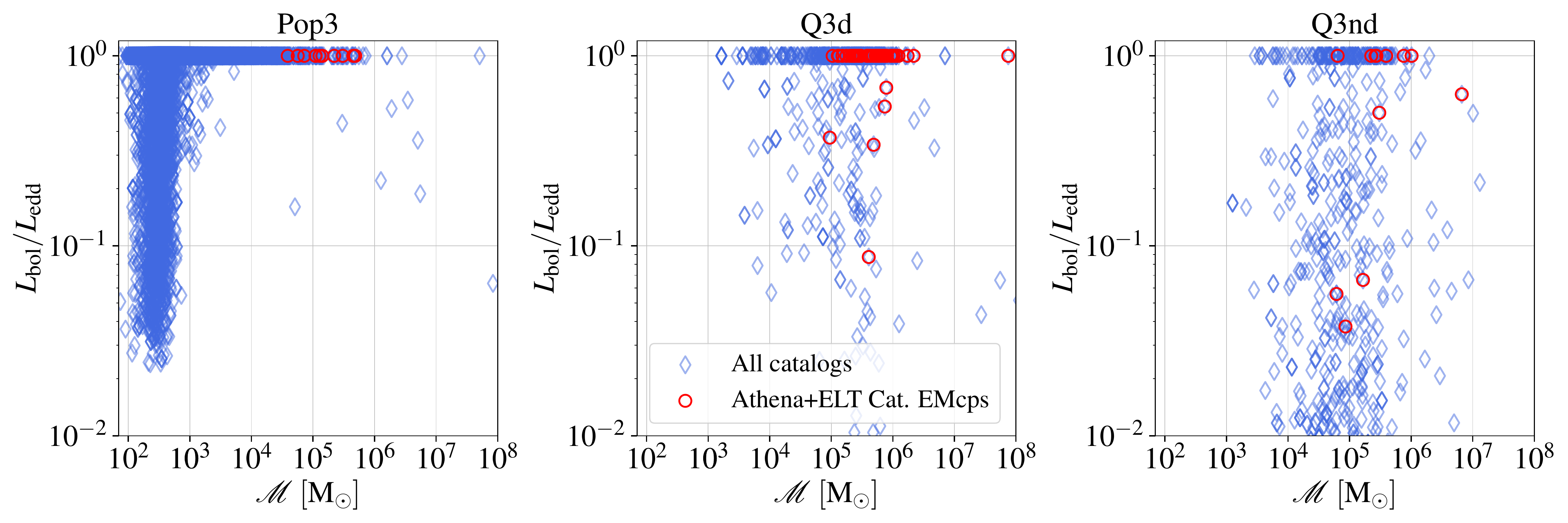} 
    \caption{Eddington ratio values as function of the chirp mass for the entire catalog (blue diamonds) and for EMcps detected by Athena+ELT in the case where the accretion is derived from the catalogues (red circles), as clarified by the legend, for the three astrophysical models.}
    \label{fig:Ledd_ratio_vs_mchirp}
\end{figure*}

\begin{figure*}
    \subfigure{\includegraphics[width=1\textwidth]{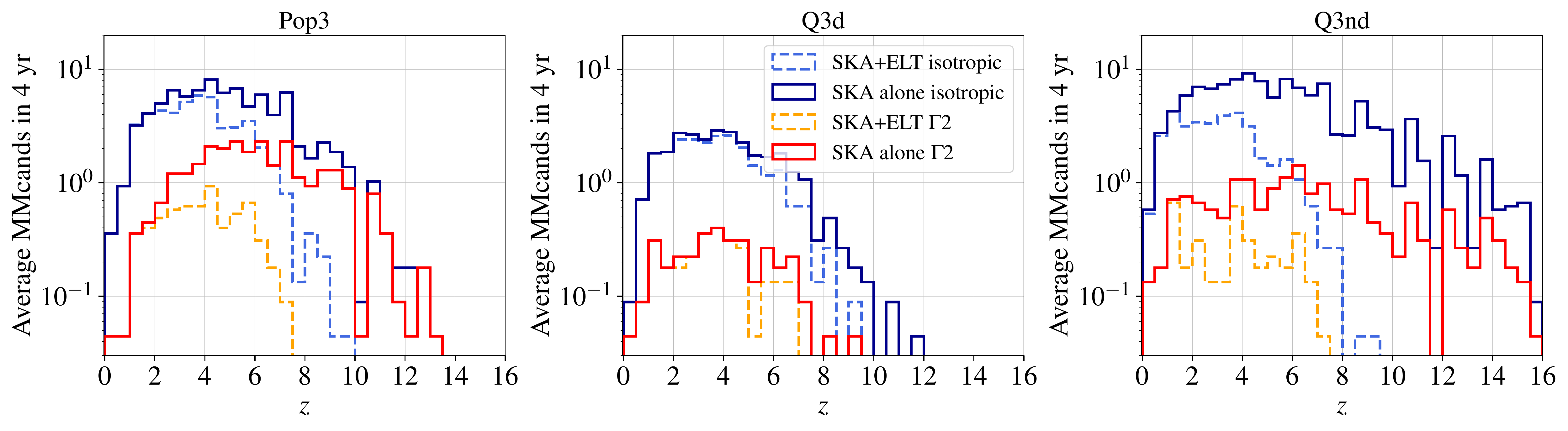}} \\ 
    \vspace{-0.6cm}
    \subfigure{\includegraphics[width=1\textwidth]{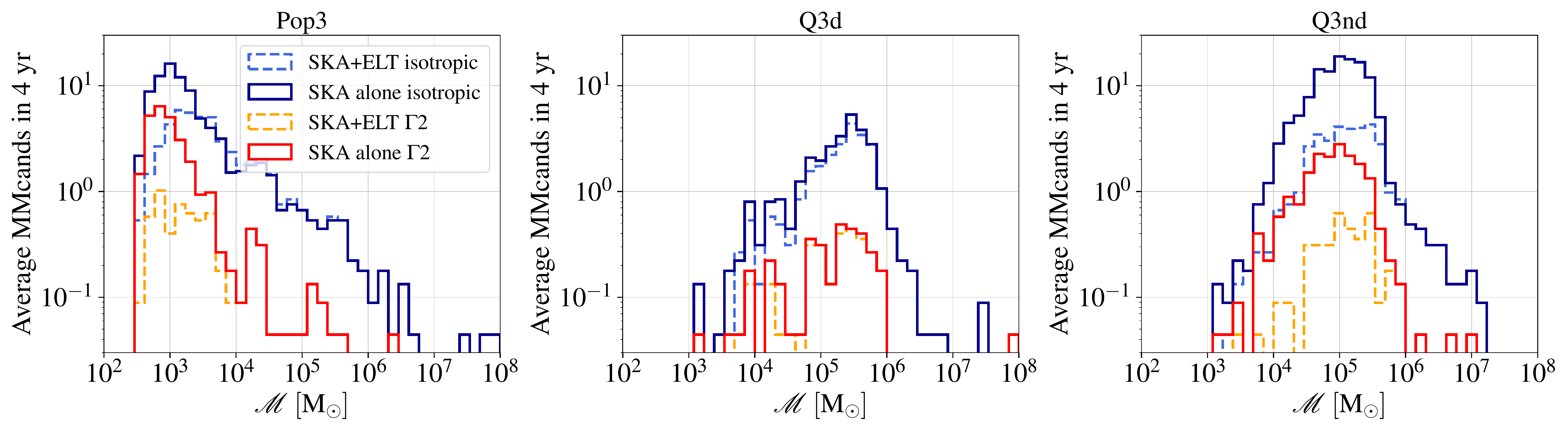}} 
    \caption{Average number of MMcands in each observational scenario, as clarified by the legend, as function of redshift (upper panels) and chirp mass (lower panels) for the three astrophysical models assuming 4 yr of LISA time mission.}
    \label{fig:mmcand_distro_ska_only}
\end{figure*}

\end{document}